\documentclass[useAMS,usenatbib]{mn2e}
\usepackage{times}
\usepackage{epsfig}
\usepackage{graphicx}
\usepackage{subfigure}
\usepackage{latexsym}
\usepackage{pifont}
\usepackage{here}
\input epsf

\newcommand{\ergs}{erg s$^{-1}$}

\newcommand{\degree}{\degr}

\def\eg{{\it e.g.,~\/}}

\def\ie{{\it i.e.,~\/}}
\def\cf{{\it cf.~\/}}
\def\deg{{${ }^{\circ}$}}

\title[The Galactic Plane at faint X-ray fluxes]{The Galactic Plane at faint 
X-ray fluxes - I:  Properties and characteristics of the X-ray source population}

\author[R.S. Warwick, D. P\'erez-Ram\'{i}rez and  K. Byckling]
{R. S. Warwick$^{1}$\thanks{E-mail:rsw@star.le.ac.uk}, 
D. P\'erez-Ram\'{i}rez$^{2}$ and  K. Byckling$^{1}$ \\
$^{1}$X-Ray \& Observational Astronomy Group, Dept. of Physics \& Astronomy,  
University of Leicester, Leicester LE1 RH7, UK\\
$^{2}$Departamento de F\'{i}sica, Universidad de Ja\'en, 
Campus Las Lagunillas, s/n, Ja\'en, E-23071, Spain}

\begin{document}

\date{Accepted . Received ; in original form }

\pagerange{\pageref{firstpage}--\pageref{lastpage}} \pubyear{2009}

\maketitle

\label{firstpage}

\begin{abstract}

We investigate the serendipitous X-ray source population revealed
in {\sl XMM-Newton} observations targeted in the Galactic Plane
within the region 315\deg $< l <$ 45\deg~and $|b|<$2.5\deg.  
Our study focuses on a sample of 2204 X-ray sources at intermediate 
to faint fluxes, which were detected in a total of 116 {\sl XMM-Newton} 
fields and are listed in the 2XMMi catalogue. We characterise each 
source as spectrally soft or hard on the basis of whether the bulk of 
the recorded counts have energies below or above 2 keV and find that 
the sample divides roughly equally (56\%:44\%) into these soft and hard 
categories. The X-ray spectral form underlying the soft sources may be 
represented as either a power-law continuum with $\Gamma \sim 2.5$ or a 
thermal spectrum with kT $\sim 0.5$ keV, with $N_H$ ranging from 
$10^{20-22} \rm~cm^{-2}$.  For the hard sources, a significantly harder 
continuum form is likely, \ie  $\Gamma \sim 1$, with $N_H = 10^{22-24} 
\rm~cm^{-2}$. For $\sim$50\% of the hard sources, the inferred column density 
is commensurate with the total Galactic line-of-sight value; many of 
these  sources will be located at significant distances across the 
Galaxy implying a hard band luminosity $L_X > 10^{32} \rm~erg~s^{-1}$, 
whereas some will be extragalactic interlopers. A high fraction 
($^{>}_{\sim}$90\%) of the soft sources have potential near-infrared 
(2MASS and/or UKIDSS) counterparts inside their error circles, consistent 
with the dominant soft X-ray source population being relatively nearby 
coronally-active stars. These stellar counterparts are generally brighter 
than J=16, a brightness cutoff which corresponds to the saturation of the 
X-ray coronal emission at $L_X = 10^{-3}~L_{bol}$. In contrast, the 
success rate in finding likely infrared counterparts to the hard X-ray 
sample is no more than $\approx15\%$ down to J=16 and $\approx25\%$ down 
to J=20, set against a rapidly rising chance coincidence rate.  The 
make-up of the hard X-ray source population, in terms of the known 
classes of accreting and non-accreting systems, remains uncertain.

\end{abstract}



\section{Introduction}
\label{sec:intro}

The brightest sources discovered in the first all-sky surveys conducted 
at X-ray wavelengths (\eg {\it Uhuru}, \citealt{Forman78}; 
{\it Ariel V}, \citealt{Warwick81}), were found to be X-ray luminous close 
binary systems powered by the accretion of matter onto a compact object.
These sources, with intrinsic X-ray luminosities (L$_X$) typically in 
the range $10^{36-38}$ \ergs~in the 2--10 keV band, can be classed either 
as low-mass or high-mass X-ray binaries (LMXBs or HMXBs) depending on the 
nature of the non-degenerate star, the companion objects being either neutron 
stars or black holes.  X-ray catalogues of the time also contained a 
few examples of other types of Galactic X-ray source, including X-ray 
bright supernova remnants, cataclysmic variables (CVs) and coronally-active 
binaries, such as RSCVn systems, albeit with inferred X-ray luminosities 
typically much less than $10^{36}$ \ergs.
 
Later more sensitive X-ray surveys utilising imaging instruments operating
in a somewhat softer spectral regime (\eg {\it Einstein}, \citealt{Hertz84}; 
{\it ROSAT}, \citealt{Voges99}) showed that the X-ray sky at low Galactic 
latitude is crowded, with nearby coronally-active stars and binaries dominating 
the source statistics in the soft ($<$ 2 keV) X-ray band (\citealt{Motch97}). 
Subsequently, an imaging survey 
in the Galactic plane carried out by {\it ASCA} (\citealt{Sugizaki01}) provided a 
first detailed view of the general population of faint Galactic X-ray sources in 
the 2--10 keV band. Since the impact of X-ray absorption by interstellar gas 
is greatly diminished above $\sim$ 2 keV, the {\it ASCA} survey was able to
detect hard-spectrum sources with X-ray luminosities markedly lower than those 
of the classical Galactic X-ray binaries across a significant fraction of the
inner Galaxy.

More recently, the improved sensitivity, spatial resolution and energy
range afforded by the {\it Chandra} and {\it XMM-Newton} observatories
has provided the opportunity to revisit the issue of the faint Galactic
source populations in both the soft and hard X-ray bands.  
For both missions, the fields-of-view of the on-board X-ray cameras are such
that in  a typical low-latitude pointing, in addition to the primary target,
many tens of serendipitous sources are seen. Both missions have also invested 
observing time in studying specific regions of the Galactic plane, most 
notably the Galactic centre, through both deep pencil-beam observations 
(\citealt{Ebisawa01}; \citealt{Ebisawa05}; \citealt{Muno03}; \citealt{Muno04};
\citealt{Hong09}; \citealt{Rev09}) and mini-surveys in which wider angle coverage 
is achieved by the mosaicing of multiple (relatively short) observations
(\citealt{Wang02}; \citealt{Hands04}; \citealt{Wijnands06};
\citealt{Muno06}; \citealt{Muno09}).  
The outcome is that 10 years post-launch both the {\it Chandra} and
{\it XMM-Newton} archives contain a wealth of data relevant to
Galactic X-ray sources at intermediate to faint flux levels,
spanning a wide range of intrinsic luminosity.

Recent results from {\it Chandra} and {\it XMM-Newton} demonstrate the 
potential of Galactic X-ray surveys, at readily accessible sensitivity limits, 
to detect a wide variety of source types. For instance, 
coronally-active binaries can be detected at distances up to 1 kpc or beyond 
(\citealt{Herent06}) and young stellar objects (YSOs) can be unveiled in 
regions of current star-formation.  In the case of the latter, evolved protostars 
and TTauri stars with extreme coronal emission and hard X-ray spectra (kT$\sim$1-4 keV) 
can be detected in dense molecular clouds despite the large line-of-sight column density
(\eg \citealt{Feigelson99}). 
Isolated Neutron Stars (ISNs), such as those discovered in the ROSAT survey (Haberl 2007), 
are radio-quiet objects located at distances of no more than a few hundred parsecs that 
display soft thermal X-ray spectra.  CVs constitute a 
source class that may, potentially, be found in large numbers in sensitive 
Galactic X-ray surveys. In particular, it has been proposed that intermediate polars 
may account for a large fraction of the low L$_{X}$ sources which reside in the Galactic 
Centre region (\citealt{Muno06}) and through their integrated emission may account
for a significant fraction of the hard Galactic X-ray Ridge emission (\citealt{Sazonov06}; 
\citealt{Rev06}). Relatively quiescent X-ray binaries with either low-mass
(\eg black hole transients) or high-mass  (\eg Be star) secondaries
may also make a non-negligible contribution to the source statistics.

However, in truth, our knowledge of the makeup of the Galactic X-ray 
source population at relatively faint levels is quite limited. This is 
certainly the case in the 2--10 keV band where, in principle, the visible 
volume extends to the edge of the Galaxy. In order to better define the 
various populations in terms of their space density, scale height and 
luminosity function,  we need to characterise and, where possible,
identify much larger samples of sources than are currently available.
More comprehensively characterised source samples might also reveal how 
X-ray sources map onto structures such as the Galactic spiral arms,
the thin and thick disc,  the Galactic bulge and the mass concentration within
$\sim$ 100 pc of the Galactic Centre.  A range of astrophysical issues, 
for example, relating to the formation and evolution of close binary systems 
and accretion at low mass-transfer rates, might also be addressed.

In the above context, the {\it Chandra} Multiwavelength Plane (ChaMPlane) 
Survey is aiming at a systematic analysis of low-latitude fields with 
the objective of measuring or constraining the populations of low-luminosity
( L$_{X}$ $_{\sim}^{>}~10^{31}$ erg s$^{-1}$) accreting white dwarfs, neutron
stars, and stellar mass black holes in the Galactic plane and bulge (see 
\citealt{Grindlay05} for full details). The programme utilises {\it Chandra} 
X-ray data in combination with follow-up optical and IR photometric and 
spectroscopic observations, so as to maximise the number of identified 
sources and explore the populations thereby revealed.  Recent publications from
the ChaMPlane programme include a study of {\it Chandra} fields in the 
Galactic anti-centre (\citealt{Hong05}) and the Galactic centre and bulge 
({\citealt{Laycock05};  \citealt{Koenig08}; \citealt{Hong09}; 
\citealt{vandenberg09}).

In the case of {\sl XMM-Newton}, the {\it XMM-Newton} Galactic Plane
Survey (hereafter XGPS; \citealt{Hands04}) has sampled a flux range 
which bridges the gap between the relatively shallow {\it ASCA} Galactic Plane 
Survey (\citealt{Sugizaki01}) and the sensitivity limits reached in deep
{\it Chandra} pointings (\eg \citealt{Ebisawa01}). In a paper reporting the
XGPS, \citet{Hands04} discussed the results from a programme including
22 pointings which cover a region of approximately 3 square degrees 
between 19\degree - 22\degree~in Galactic longitude and $\pm$0.6\degree~
in latitude.  Subsequent optical follow-up observations of a representative
sample of the brightest identified low-latitude hard band sources
detected in the XGPS have recently been presented by \citet{Motch10}.

In the present paper, we build on the XGPS studies of \citet{Hands04} 
by carrying out a systematic investigation of the X-ray source population
seen at intermediate to faint fluxes in {\it XMM-Newton} pointings targeted 
at the Galactic plane.  More specifically, we consider observations  
encompassing a narrow strip of the Plane towards the central quadrant of 
the Galaxy. In the next section of this paper, we give details of the 
set of {\sl XMM-Newton} observations which 
provide the basis of the study. 
In Section~3, we describe the selection criteria we employ to extract a 
clean ``serendipitous'' X-ray source sample from these fields, using
the 2XMMi catalogue (\citealt{Watson09}) as the input database.
Section 4 goes on to investigate various properties of the sample
with a focus on the available X-ray spectral information. 
In section 5 we present an investigation of the likely near-infrared
counterparts based on a cross-correlation of the X-ray sample with both 
the 2MASS and UKIDSS surveys.  This leads on to a discussion of the
nature of the soft- and hard-source populations in section 6,
followed by a brief summary of our conclusions of this first paper (Paper I).  
In subsequent papers (in preparation), we will explore the average X-ray spectral 
properties of a subset of relatively bright sources drawn from our source sample 
(Paper II), the log N - log S curves for both the soft and hard source samples
(Paper III), and the broad-band colours of likely counterparts to the X-ray
sources (Paper IV).


\section{The {\sl XMM-Newton} Galactic Plane Database}
\label{sec:survey}

In this paper we utilise observations drawn from the {\sl XMM-Newton}
public data archive targeted at positions along the Galactic plane 
within the central quadrant of the Galaxy. More specifically we
use observations with pointings in the region bounded by 315\deg $< l <$ 45\deg~ 
and $|b|<$2.5\deg. Our preliminary list of observations comprised 
those used in constructing the 2XMMi catalogue (\citealt{Watson09}) 
- see Fig. \ref{fig:all_fields}. However, given our focus on the 
serendipitous source content of the {\sl XMM-Newton} fields, we excluded
those fields dominated by a very bright source, which in most cases was 
the target source. We also excluded a number of observations otherwise 
dominated by the target, such as those containing nearby star clusters or 
bright extended emission arising from supernova remnants.  In some instances 
we excised the region within the field of view dominated by the bright 
emission but retained the rest of the observation.  Observations 
which significantly overlapped with other deeper observations were also 
removed, although fields with fractional overlaps giving mosaiced coverage 
of particular regions were retained.  A couple of observations with relatively 
short net exposures affording only poor sensitivity were also discarded.  

This filtering process resulted in a final list of 116 observations
which comprise the present survey. Details of the set of selected
observations are given in Table \ref{tab:obs_info} and the corresponding
distribution of pointing directions along the Galactic plane 
is illustrated in Fig. \ref{fig:all_fields}.  In brief, 
Table \ref{tab:obs_info} provides the following information
for each field: {\it column 1} - the {\sl XMM-Newton} observation ID;
{\it columns 2-3} - the nominal observation pointing position
in Galactic coordinates ({\it l,b}); {\it columns 4-6} - net exposure
time in the three EPIC cameras;  {\it columns 7-9} - the filter
deployed in each EPIC camera; {\it column 10} - the fraction of the
total field of view retained in those fields in which a sub-region 
was excluded; {\it columns 11-13} - the number of sources 
detected in the field which pass the selection criteria 
defined in \S 3 (both the total number and the split between 
the soft and hard assignments are given).

Most of the observations were performed with the three EPIC cameras 
(pn, MOS-1 and MOS-2; \citealt{Strud01}; \citealt{Turner01})
in operation in a {\it Full-Frame Window} mode for which all the pixels of 
all the CCDs are read out and the full field-of-view is covered.
For five of the observations no suitable pn data were available. Although, 
in most cases, the Medium filter was selected, there were a few occasions 
when either the Thin1 or Thick filters were deployed 
(see Table \ref{tab:obs_info}).  
The distribution of the effective exposure time, calculated as the
average of the individual pn, MOS1 and MOS2 exposures, with a double 
weighting for the more sensitive pn instrument, is illustrated in Fig.
\ref{fig:exp_times}; the minimum effective exposure was
1.5 ks and the maximum 71.3 ks.  The total sky area covered by the 
set of survey fields was 22.5 sq.deg.


\begin{figure}
\begin{center}
{\scalebox{0.55}{\includegraphics{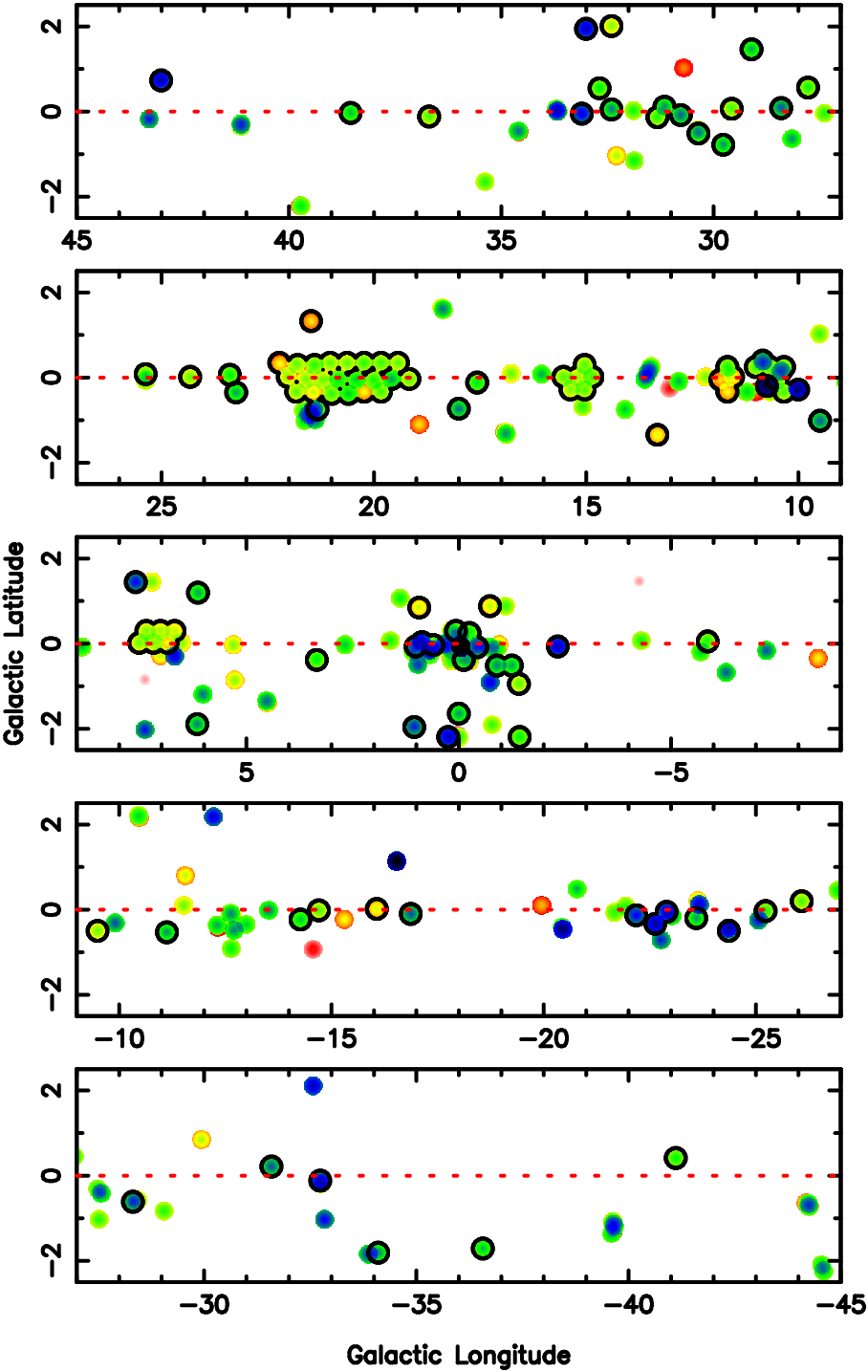}}}
\caption{The spatial distribution of the {\sl XMM-Newton} 
observations that fall within the Galactic plane region
bounded by $|l|<$ 45\deg and $|b|<$2.5\deg. Each colour dot 
corresponds to a field used in the construction of the 2XMMi 
catalogue  (\citealt{Watson09}), with the colour coding
red-to-green-to-blue roughly scaling as the logarithm of the
accumulated exposure. The 116 observations included 
in the present study are circled in black. 
}
\label{fig:all_fields}
\end{center}
\end{figure}


\begin{figure}
\begin{center}
{\includegraphics[width=4.5cm,angle=270]{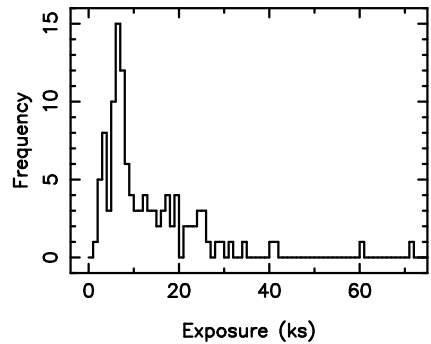}}
\caption{The distribution of the effective exposure time across the 
set of 116 observations, calculated as the
average of the individual pn, MOS1 and MOS2 exposures with a double 
weighting for the more sensitive pn instrument.}
\label{fig:exp_times}
\end{center}
\end{figure}


\begin{table*}
  \caption{Details of the 116 {\sl XMM-Newton} observations which 
comprise the current Galactic Plane survey}
  \begin{center}
  \begin{tabular}{lcccccccccccc}
  \hline
   Observation & \multicolumn{2}{c}{Pointing}   & \multicolumn{3}{c}{Exposure (ks)} 
& \multicolumn{3}{c} {Filter} &  FoV  & \multicolumn{3}{c} {Sources}\\
   ID & $l$ & $b$ & pn & MOS1 & MOS2 & pn & MOS1 & MOS2 & Fraction & total & soft & hard\\
 \hline

0400910201&318.9&$+$0.5& 9.5&11.6&11.6&Med&Med&Med&   &13 &12 &1  \\
0405390401&323.4&$-$1.7&12.4&16.6&16.3&Med&Med&Med& 0.97 &13 &8 &5  \\
0400890101&325.9&$-$1.8&15.4&21.6&23.2&Thin1&Thin1&Thin1& 0.88  &27 &19 &8  \\
0203910101&327.2&$-$0.1& 7.1& 8.7&8.7&Med&Med&Thin1&0.95&16 &10 &6 \\
0143380101&328.4&$+$0.2& 0.0&50.0&50.2&  &Thick&Thick&0.98&34 &26 &8 \\
0406650101&331.7&$-$0.6&24.7&31.6&31.6&Med&Med&Thin1& &61 &47 &14 \\
0406750201&333.8&$+$0.1&6.0&7.6&7.6&Med&Med&Med& &18 &13 &5 \\
0403280201&334.7&$+$0.0&7.5&9.9&11.1&Med&Med&Med&0.82 &23 &16 &7 \\
0201000401&335.6&$-$0.4&18.1&21.0&21.3&Thin1&Thin1&Thin1&0.98&32 &20 &12\\
0022140101&336.4&$-$0.2& 4.3&13.8&14.7&Med&Med&Med&0.97& 8& 5 &3 \\
0204500301&337.1&$+$0.0&30.7&32.6&32.6&Thin1&Thin1&Thin1&0.88&28 &15 &13 \\
0307170201&337.4&$-$0.3&60.0&81.2&84.0&Med&Med&Med& &70 &48 &22 \\
0303100101&337.8&$-$0.1&38.3&45.3&45.7&Med&Med&Med&0.97&51 &37 &14 \\
0200900101&343.1&$-$0.1&23.9&27.3&27.6&Med&Med&Med& &28 &15 &13  \\
0406752301&343.9&$+$0.0& 2.7& 3.5&4.3&Med&Med&Med& &9 &3 &6  \\
0112460201&345.0&$+$2.5&20.9&27.7&27.9&Med&Med&Med& &26 &12 &14  \\
0406750301&345.3&$+$0.0& 6.1& 7.6& 7.6&Med&Med&Med& &18 &13 &5  \\
0144080101&345.7&$-$0.2&10.9&16.9&16.9&Med&Med&Med& &28 &18 &10  \\
0401960101&348.9&$-$0.5&15.5&19.2&20.3&Med&Med&Med& &28 &18 &10  \\
0103261201&350.5&$-$0.5& 4.3& 6.7& 6.7&Med&Med&Med& &8 &5 &3  \\
0112201401&354.1&$+$0.1& 4.6& 6.3& 6.4&Med&Med&Med&0.77&7 & 3&4 \\
0301730101&357.6&$-$0.1&59.8&61.5&61.5&Thick&Thick&Thick&0.95&67 & 40&27 \\
0103261301&358.6&$-$1.0& 5.0& 7.6& 7.6&Med&Med&Med& &10 &2 &8  \\
0206990101&358.6&$-$2.2& 7.9&13.5&13.5&Med&Med&Med& &43 &28 &15 \\
0406580201&358.8&$-$0.5&14.1&18.4&19.1&Med&Med&Med& &21 &9 &12 \\
0400340101&359.1&$-$0.5&17.2&20.2&22.3&Med&Med&Med&0.95&25 &10 &15 \\
0304220101&359.3&$+$0.9& 0.0& 7.5& 7.7&   &Med&Med&0.98 &5 &4 &1  \\
0112970701&359.6&$+$0.0&19.4&23.4&23.4&Med&Med&Med& &59 &13 &46 \\
0302884301&359.7&$+$0.3& 5.0& 6.5& 6.5&Med&Med&Med& &19 &6 &13 \\
0112971001&359.9&$-$0.3&11.3&11.3&11.5&Thick&Med&Med& &26 &15 &11  \\
0112971901&  0.0&$+$0.3& 4.2& 7.5& 7.2&Med&Med&Med& & 11& 3& 8 \\
0307110101&  0.0&$-$1.6&13.3&21.2&21.9&Med&Med&Med& &24 &14 &10  \\
0206590201&  0.3&$-$2.2&17.0&20.5&20.5&Med&Med&Med& &42 &29 &13  \\
0112971501&  0.6&$+$0.0& 6.9& 9.3& 9.3&Med&Med&Med&  &14 &7 &7  \\
0112970201&  0.9&$+$0.1&12.9&17.4&17.4&Med&Med&Med&0.98&38 &15 &23 \\
0305830701&  0.9&$+$0.8& 3.0& 4.0& 4.4&Med&Med&Med&  & 3 & 1 & 2 \\
0205240101&  1.0&$-$0.1&15.1&32.8&34.3&Med&Med&Med&  &40 &16 &24 \\
0302570101&  1.0&$-$1.9&27.1&31.1&31.1&Med&Med&Med&  & 60&34 &26\\
0148090501&  3.3&$-$0.3& 6.5& 8.8& 8.7&Thin1&Thin1&Thin1&  &16 &8 &8 \\
0301881001&  6.1&$+$1.2&14.1&15.9&14.8&Med&Med&Med&  & 17& 9&8 \\
0301540101&  6.1&$-$1.9&18.2&21.3&21.2&Thin1&Thin1&Thin1&  &36 &21 &15 \\
0135742301$^\dag$&  6.7&$+$0.3& 3.5& 4.1& 4.2&Med&Med&Med&0.81& 4&3 &1 \\
0135742501$^\dag$&  6.8&$+$0.0& 2.3& 2.8& 2.9&Med&Med&Med&0.71& 3&2 &1 \\
0135742601$^\dag$&  7.0&$+$0.3& 4.0& 4.9& 4.9&Med&Med&Med& &10  &8 &2  \\
0135742801$^\dag$&  7.2&$+$0.0& 4.7& 5.8& 6.0&Med&Med&Med&  &17 &16 &1  \\
0135742901$^\dag$&  7.3&$+$0.3& 4.4& 6.1& 6.3&Med&Med&Med& & 12 &10 &2  \\
0135743101$^\dag$&  7.5&$+$0.0& 4.9& 6.6& 6.6&Med&Med&Med&  & 9& 7& 2 \\
0099760201&  7.6&$+$1.4&39.3&41.6&41.4&Med&Med&Med&  &55 &22 &33 \\
0145840201&  9.4&$-$1.0&19.7&25.4&25.7&Thin1&Thin1&Thin1&  &54 &38 &16 \\
0164561401& 10.0&$-$0.2& 0.0&32.0&32.6&   &Med&Med&0.92&21 &13 &8  \\
0152832801$^\dag$& 10.3&$+$0.3& 7.6& 9.4& 9.4&Med&Med&Med&  &13 &8 &5\\
0152832901$^\dag$& 10.3&$-$0.3& 5.6& 8.5& 8.5&Med&Med&Med&  & 9&6 &3 \\
0145840101& 10.4&$+$0.1&20.2&26.2&26.5&Thin1&Thin1&Thin1&  & 33&18 &15 \\
0152833001$^\dag$ &10.5&$+$0.0& 5.7& 8.2& 8.3&Med&Med&Med&0.93& 13& 10& 3 \\
0301270401&10.7&$-$0.2& 29.3&39.0& 22.0&Med&Med&Med& 0.96 & 46& 32& 14 \\
0152833101$^\dag$& 10.7&$+$0.3& 7.0& 8.7& 8.7&Med&Med&Med& & 9& 7&2  \\
0024940201& 10.8&$+$0.4&22.5&25.2&25.2&Med&Med&Med&0.98&50 &33 &17 \\
0152833401$^\dag$& 11.0&$+$0.3& 7.0& 8.7& 8.7&Med&Med&Med& &22 &17 &5  \\
0152835101$^\dag$& 11.5&$+$0.0& 1.4& 4.8& 5.5&Med&Med&Med& &3 &2 &1  \\
0152835501$^\dag$& 11.7&$-$0.3& 1.7& 3.6& 3.5&Med&Med&Med& &4 &3 &1  \\
0152835701$^\dag$& 11.7&$+$0.3& 8.0&10.7&10.7&Med&Med&Med& &17 &10 &7  \\

\hline
\end{tabular}
\end{center}
\end{table*}

\setcounter{table}{0}

\begin{table*}
  \caption{Continued}
  \begin{center}
  \begin{tabular}{lcccccccccccc}
  \hline
   Observation & \multicolumn{2}{c}{Pointings}   & \multicolumn{3}{c}{Exposure (
ks)} & \multicolumn{3}{c} {Filter} &  FoV & \multicolumn{3}{c} {Sources}\\ 
   ID & $l$ & $b$ & pn & MOS1 & MOS2 & pn & MOS1 & MOS2 & Fraction & total & soft & hard\\
 \hline

0152835401$^\dag$& 11.8&$+$0.0& 1.2& 3.0& 3.2&Med&Med&Med& & 3 &1 &2  \\
0150960301& 13.3&$-$1.3& 1.1& 4.8& 5.0&Med&Med&Med& &3 &3 &0  \\
0152835001$^\dag$& 14.8&$+$0.0& 6.9& 8.5& 8.5&Med&Med&Med& &20 &11 &9  \\
0152834401$^\dag$& 15.0&$+$0.3& 4.2& 8.7& 9.5&Med&Med&Med&  &11&8 &3  \\
0152834501$^\dag$& 15.0&$-$0.3& 4.2& 8.9& 9.5&Med&Med&Med&  &11&4 &7  \\
0152834601$^\dag$& 15.2&$+$0.0& 3.7& 6.1& 6.3&Med&Med&Med& & 7& 2 &5 \\
0152834801$^\dag$& 15.3&$-$0.3& 3.7& 7.0& 7.1&Med&Med&Med& & 8& 6& 2 \\
0152834901$^\dag$& 15.5&$+$0.0& 5.0& 8.4& 8.4&Med&Med&Med& & 18& 9 &9  \\
0040140201& 17.5&$-$0.1&12.3&11.3&12.3&Med&Med&Med& &18 &7 &11  \\
0054540701& 18.0&$-$0.7& 0.0&23.1&23.0&   &Med&Med&0.98&15 &6 &9 \\
0104460701& 19.2&$+$0.0& 5.4& 8.4& 7.6&Med&Med&Med&  &11 &6 &5 \\
0051940101$^\dag$& 19.4&$+$0.3& 4.8& 7.8& 7.8&Med&Med&Med&  &10 &5 &5 \\
0135745401$^\dag$& 19.6&$+$0.0&11.3&13.7&13.7&Med&Med&Med&  &13 &6 &7 \\
0051940301$^\dag$& 19.8&$+$0.3& 6.1& 9.1& 9.1&Med&Med&Med&  & 9& 3& 6\\
0051940401$^\dag$& 19.8&$-$0.3& 5.7& 9.1& 9.1&Med&Med&Med&  &16 &10 &6\\
0104460301& 20.0&$+$0.0& 7.7&11.2&11.0&Med&Med&Med&  & 18& 10&8\\
0051940601$^\dag$& 20.2&$-$0.3& 1.4& 3.8& 3.8&Med&Med&Med&  & 2& 1&1\\
0051940501$^\dag$& 20.2&$+$0.3& 6.8& 9.6& 9.6&Med&Med&Med&  & 10&4 &6\\
0104460401& 20.4&$+$0.0& 5.3&10.3&10.0&Med&Med&Med&  & 9&1 &8\\
0135740701$^\dag$& 20.6&$+$0.3& 4.8& 7.3& 7.4&Med&Med&Med&  & 8&2 &6\\
0135745801$^\dag$& 20.6&$-$0.3&11.5&13.2&13.2&Med&Med&Med&  & 15&3 &12\\
0104460901& 20.8&$+$0.0&11.3&13.6&13.6&Med&Med&Med&  & 16&3 &13\\
0135740901$^\dag$& 21.0&$+$0.3& 4.8& 7.8& 7.7&Med&Med&Med&  & 11&5&6\\
0135745901$^\dag$& 21.0&$-$0.3& 5.8& 9.1& 9.5&Med&Med&Med&  & 7& 2&5\\
0135745201$^\dag$& 21.2&$+$0.0& 4.8& 7.1& 7.3&Med&Med&Med&  & 7&5 &2\\
0122700801& 21.3&$-$0.8&12.5&15.0&14.8&Med&Med&Med&0.94& 14&5 &9\\
0112201001& 21.4&$+$1.3& 0.0& 4.4& 4.5&   &Med&Med&  &4 &2 &2\\
0135745701$^\dag$& 21.4&$+$0.3& 5.7& 7.4& 7.4&Med&Med&Med&  &10 &5 &5\\
0135746801$^\dag$& 21.4&$-$0.3& 2.7& 4.4& 4.4&Med&Med&Med&  &7 &2 &5\\
0135746301$^\dag$& 21.6&$+$0.0& 4.3& 6.7& 6.7&Med&Med&Med&  &6 &2 &4\\
0135746401$^\dag$& 21.8&$+$0.3& 1.7& 4.3& 4.5&Med&Med&Med&  &5 &1 &4\\
0135741601$^\dag$& 21.8&$-$0.3& 4.7& 7.7& 7.7&Med&Med&Med&  &11 &1 &10\\
0135741701$^\dag$& 22.0&$+$0.0& 3.6& 6.5& 6.8&Med&Med&Med&  &6 & 0&6\\
0135744801$^\dag$& 22.2&$+$0.3& 1.1& 1.8& 2.0&Med&Med&Med&  &2 &1 &1\\
0302560301& 23.2&$-$0.3&15.1&17.9&18.2&Med&Med&Med&  &22 &9 &13\\
0400910101& 23.4&$+$0.1&10.2&12.1&12.1&Med&Med&Med&  &11 &3 &8\\
0203850201& 24.3&$+$0.1& 5.3& 6.8& 7.2&Med&Med&Med&0.89&10 &6 &4\\
0400910301& 25.4&$+$0.1& 7.5& 9.1& 9.1&Med&Med&Med& &20 &11 &9\\
0301880401& 27.7&$+$0.6& 8.4&10.2&10.1&Med&Med&Med&  & 6& 3&3\\
0302970301& 28.4&$+$0.1&22.7&26.7&26.7&Med&Med&Med&  &28& 18&10\\
0301880901& 29.1&$+$1.5&14.9&16.6&16.6&Med&Med&Med&0.88&17 &7 &10\\
0046540201& 29.5&$+$0.1& 6.7& 8.0& 8.4&Med&Med&Med& & 7& 3 & 4\\
0406140201& 29.8&$-$0.7&17.0&18.5&18.5&Med&Med&Med&0.98 & 15& 4 &11\\
0302970801& 30.3&$-$0.5&11.7&15.5&15.6&Med&Med&Med& & 28&21 &7\\
0203850101& 30.8&$+$0.0&20.6&25.2&25.2&Med&Med&Med&0.97& 31&18 &13\\
0203910201& 31.3&$-$0.1& 5.4&11.8&12.2&Med&Med&Med&0.85&6 &6 &0\\
0207010201& 31.1&$+$0.2&17.4&18.7&18.7&Med&Med&Med&0.89&7 &1 &6\\
0211080101& 32.4&$+$2.1& 3.3& 6.8& 6.6&Thin1&Thin1&Thin1&  & 7&6 &1 \\
0136030101& 32.4&$+$0.1&17.8&21.4&21.9&Med&Med&Med&0.90&27 &15 &12\\
0306170201& 32.7&$+$0.5& 9.9&11.5&11.5&Thin1&Med&Med&0.98 &7 &4 &3\\
0017740601& 33.0&$+$2.0&20.0&22.3&22.3&Med&Med&Med&  & 34&15 &19\\
0017740401& 33.1&$+$0.0&25.7&27.7&27.7&Med&Med&Med&0.77&16 &8 &8\\
0302970201& 36.7&$-$0.1&2.7&10.0&13.3&Med&Med&Med& &6 &3 &3\\
0136030201& 38.5&$+$0.0&13.3&16.0&16.0&Med&Med&Med&0.98&10 &6 &4\\
0305580201& 43.0&$+$0.8&22.3&27.8&28.2&Med&Med&Med&0.89& 22& 6&16\\

\hline
\end{tabular}
\end{center}
{\small $^\dag$ Observations corresponding to the Galactic X-ray 
Survey Programme (XGPS)}
\label{tab:obs_info}
\end{table*}



\section{Defining a serendipitous source sample}
\label{sec:sample}

As the starting point for the compilation of a serendipitous
X-ray source sample, we extracted a complete list of the sources detected 
in the 116 designated fields  from the Second {\it XMM-Newton} 
Serendipitous Source Catalogue (2XMMi; \citealt{Watson09}).  
The 2XMMi catalogue provides extensive details of the source properties 
including the net count rate and detection maximum likelihood of 
each source in five separate energy bands: Band~1 (0.2-0.5 keV); 
Band~2 (0.5-1.0 keV); Band~3 (1.0-2.0 keV); Band~4 (2.0-4.5 keV): 
and Band~5 (4.5-12.0 keV). This information is provided for each of 
the three EPIC instruments.

A preliminary step involved dealing with the relatively small number of 
observations in which instrument filters other than the medium filter were 
deployed. In these cases the recorded count rates measured in each camera and 
in each energy band were scaled to equivalent values for the medium filter,
where the scale factors applied were derived from the ratio of the
appropriate energy conversion factors (ECF) quoted in \cite{Mateos09}
\footnote{These ECFs are calculated assuming a broad-band source spectrum, 
characterised as a power-law continuum with photon index $\Gamma$ = 1.7 
absorbed by a line-of-sight column density 
$\textit{N}_H$ = 3$\times$10$^{20}$ cm$^{-2}$.}. 

In our study of intermediate to faint Galactic Plane sources, it has proven 
convenient to compress the available spectral information into two bands, 
a {\sl soft band} representing the combination of Bands 2 and 3,
encompassing the energy range 0.5--2 keV, and a {\sl hard band} based on
Bands 4 and 5, corresponding to 2--12 keV.
The 2XMMi~~Band~1 information has not been used in this analysis. 
The count rates measured in each EPIC camera for the soft and hard 
bands were obtained as the straight sum of the rates recorded in the 
respective input bands.  We also calculated an effective detection 
maximum likelihood ({\it maxl}) in the soft and hard bands by the prescription 
given in the SAS {\it emldetect} documentation (see also \citealt{Mateos08}).

A further compression of the information for each detected source was 
achieved by combining the count rates measured in the different cameras 
for the soft band and, separately, for the hard band.  For the count-rate 
measurements for a particular source in a given camera to be considered 
valid,  a requirement was that the 2XMMi source parameter,  {\sl frac}, 
should be greater than $0.8$ for that camera (\ie that at least $80\%$ 
of the source response was contained within the camera's active field-of-view). 
Count-rate measurements passing this criterion were included irrespective of 
whether the source was classed as detected or not in the camera/band 
combination under consideration (see below). The actual compression process 
involved first averaging valid MOS-1 and MOS-2 count rates using a 
statistical weighting to give a combined-MOS count rate estimate and then 
scaling the latter measurement to pn equivalent units\footnote{The scale factor for
converting the combined-MOS to pn count rates was determined for 
each band on the basis of the average pn:combined-MOS count rate ratio 
derived for the full source sample}. Finally the scaled-MOS and pn
measurements were averaged, again using a statistical weighting, to
give the final soft- and hard-band count rates.
In following this process, the resultant source count rates
may derive from one, two or three measurements depending on which
combination of cameras were producing valid data for the source. 
Hereafter in this paper we quote count rates in the soft, hard 
and total (soft$+$hard) bands for the combined EPIC cameras 
in units of pn count ks$^{-1}$.

The first step in excising relatively low-significance sources from the full 
2XMMi source list involved the setting of a threshold value of {\it maxl} = 10 for the 
soft/hard band detection.
Sources were only retained in the source list if this threshold was exceeded 
in either the soft or hard bands (or both) and in one or more of the EPIC 
cameras. For a source detection to be classed as {\sl valid} in a particular 
camera, we required {\sl frac} $>0.8$ in that camera. Further criteria for source 
removal included  the 2XMMi parameter {\sl Extent} $>$ 0 (\ie
the exclusion of all non-pointlike sources) and the {\sl SUM-FLAG}
$>$ 3 (\ie a quality check to exclude sources described as ``located in an area where 
spurious detections may occur and may possibly be spurious'' - see \citealt{Watson09}). 

After filtering the original 2XMMi source list using the above 
methodology, we visually inspected the soft- and hard-band images 
of each field (produced from  the standard data products 
available from the {\sl XMM-Newton} archive).  As a result a number of sources 
classified at this stage as  ``valid detections'' were flagged for 
removal. Frequently such source exclusions stemmed from the confusion 
of one source with another, generally brighter, object.  A few sources
embedded in bright diffuse regions were also removed. At this point we also 
excluded regions encompassing an obvious bright target source - essentially by 
spatially masking areas within the observation field of view. 
On occasion, regions affected by bright extended objects or by the scattering 
rings from very bright out-of-field-of-view sources were also excluded.  
Observations in which the target region or other specific regions 
have been excluded are flagged in Table 1. 

Implementation of the above selection process left us with a catalogue of 
$\sim$ 2700 sources. A source could be included in this list based on 
a nominal detection (above a maximum likelihood threshold of 10) in just 
one band and one camera. Visual inspection suggested that at the limit, 
such sources were typically quite faint and in some cases potentially 
unreliable. Since a detailed analysis of the reliability of 2XMMi sources 
as a function of the detection maximum likelihood threshold was beyond the 
scope of the investigation (but see \citealt{Watson09}; \citealt{Mateos09}), 
we instead employed an empirical approach to arrive at a robust final source 
catalogue based on a further {\it signal-to-noise} selection. More 
specifically, we calculated the ratio of the count rate to count-rate error 
for each source using the {\it combined camera} measurements.
We then set a minimum requirement that this ratio should exceed
a value of 4 in the soft and/or hard bands. This further 
selection resulted in the exclusion of 252 sources detected 
only in the pn camera and 101 (133) sources seen only in the MOS-1 (MOS-2)
cameras, with just 10 sources removed for which there was a simultaneous 
detection in more than one camera. 
 
After applying all the above, we were left with a catalogue of 
2204 point sources drawn from the 116 survey fields. Table \ref{stats}
summarises the source detection statistics in terms of 
the soft/hard bands and the pn/MOS-1/MOS-2 cameras.

Although the field selection described in \S2 involved the exclusion 
of duplicate observations of the same target or pairs of observations 
with fields-of-view which significantly intersect,  some modest degree 
of field overlap was permitted.  As a consequence of this overlap,
there are a total 44 source duplications in the total of 2204, 
\ie 2\% of the sample.  Within these 44 sources, 30 of which are soft
detections and 14 hard, there are two instances of the same source 
being twice duplicated. In the following analysis we have ignored 
this duplication and treated all 2204 entries as individual 
sources.


\begin{table}
\caption{Source detection statistics.
}
\centering
\begin{tabular}{lrrrr}
\hline
Camera      &  Soft-Band       & Hard-Band &  Both Soft &   All \\
            &  Only            & Only      &  and Hard  &      \\  
\hline
pn only$^{a}$       &  349     &  222       &  18       & 589  \\
MOS-1 only$^{a}$    &   55     & 57        &  3         & 115  \\
MOS-2 only$^{a}$    &   56     & 67        &  7         & 130  \\
Multiple$^{b}$      &  626     & 453       &  291       & 1370 \\
\hline
Total               &  1086     & 799       &  319       & 2204 \\

\hline
\end{tabular}
\\
$^{a}$ Detections in a single camera only \\
$^{b}$ Detections in at least two cameras \\
\label{stats}
\end{table}


\section{Properties of the Source Sample}
\subsection{Spatial distribution}
\label{sec:field_dist}

The distribution in Galactic longitude and latitude of the sources
which comprise our serendipitous source sample is 
illustrated in Fig \ref{fig:latlong}. These are the observed 
distributions without any correction applied for sky coverage 
or for the depth of the underlying observations. 

It is evident from Figs. \ref{fig:all_fields} and \ref{fig:latlong}
that the coverage of the Galactic plane by {\sl XMM-Newton} is far 
from uniform.  Near the Galactic Centre, within $\left|l\right|$$<$2\deg, 
there is an obvious concentration of pointings.
Similarly the regions near l = 7\deg, 11\deg, 15\deg~and
between 19\deg-22\deg~have relative good coverage in narrow strips
along the Plane.  The latter corresponds to the ``XGPS region'' 
studied by \citet{Hands04}, whereas the other three directions have
enhanced coverage as a result of later phases of the XGPS programme. 
The observations drawn from the XGPS programme are flagged as such in 
Table \ref{tab:obs_info} - generally these consist of sequences of 
relatively short exposures with some overlapping of the fields of view so 
as to give extended, albeit shallow, sky coverage.

\begin{figure}
\centering
\includegraphics[width=5.0cm,angle=270]{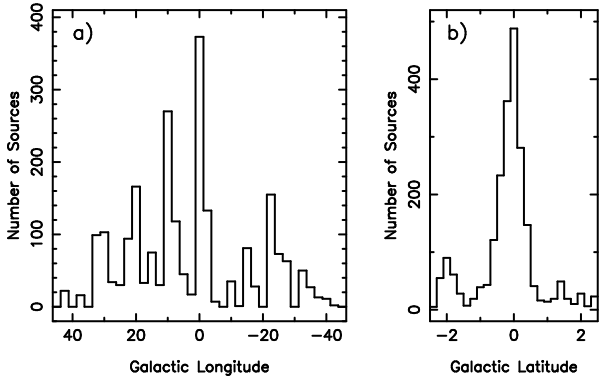}
\caption{ a) The source distribution in Galactic longitude. No 
corrections have been applied for the number or depth of the 
observations in the survey; b) The same in Galactic latitude.}
     \label{fig:latlong}
\end{figure}



\subsection{Flux calibration and distribution}
\label{sec:flux_range}
The distribution of measured {\sl total} count rate for the source
sample is illustrated in Fig. \ref{fig:countrates}.  Although the full 
distribution extends over 3 decades, $94\%$ 
of the sources have total count rates  in the range 2-100 pn ct ks$^{-1}$.  
Figure \ref{fig:hrcount} shows how the total count rate splits between the 
soft and hard bands. Roughly 50\% of the sources are detected 
only in the soft band, with 36\% detected only in the hard band and 
just 14\% detected in both bands (see Table \ref{stats}).  With the
objective of dividing the sample into non-overlapping soft and hard 
subsets, we calculate a broad-band hardness ratio,
$\rm HR = (H-S)/(H+S)$, where S and H refer to the soft-band 
and hard-band count rates respectively.  We then use HR=0, 
corresponding to the diagonal line in Fig. \ref{fig:hrcount},
(\ie equal soft and hard count rates), as the boundary between 
the two designations. This provides a close match to the soft-only 
and hard-only detections; on the other hand, for the sources detected 
in both bands, the division somewhat favours the hard category.  
Hereafter we refer to sources as {\sl soft} or {\sl hard} depending on their
position relative to the $\rm HR = 0$ fiducial. On this basis there are
1227 soft sources and 977 hard sources in our sample.

Given the very different characteristics of the underlying source populations 
(see below), different spectral forms have been assumed for the soft and 
hard sources in deriving factors to convert the measured count rates to 
energy flux.  In both bands we apply a spectral model consisting of 
a power-law continuum of photon index
$\Gamma$ subject to line-of-sight absorption in a column
density, $\textit{N}_H$.  For the soft sources, the spectral 
parameters  were set to $\Gamma$ = 2.5 and 
$\textit{N}_H$ = 10$^{21}$ cm$^{-2}$ 
resulting in an energy conversion factor (ECF) of 4.9$\times$10$^{11}$ ct 
cm$^{2}$ erg$^{-1}$,  where both the count rate and {\it unabsorbed} 
energy flux relate to 
the 0.5--2 keV band.  For the hard sources, we use
$\Gamma$ = 1.0 and $\textit{N}_H$ = 3$\times$10$^{22}$ cm$^{-2}$ leading
to an ECF of 7.9$\times$10$^{10}$ ct cm$^{2}$ erg$^{-1}$, applicable to
the 2--10 keV band.
For a source with a nominal count rate of 2 pn ct ks$^{-1}$
in the soft band, the equivalent unabsorbed
flux is  4$\times$10$^{-15}$ erg s$^{-1}$ cm$^{-2}$ (0.5--2 keV).
The same count rate in the hard band equates to an unabsorbed flux of
2.5$\times$10$^{-14}$ erg s$^{-1}$ cm$^{-2}$ (2--10 keV).


\begin{figure}

\centering
\rotatebox{-90}{\scalebox{0.30}{\includegraphics{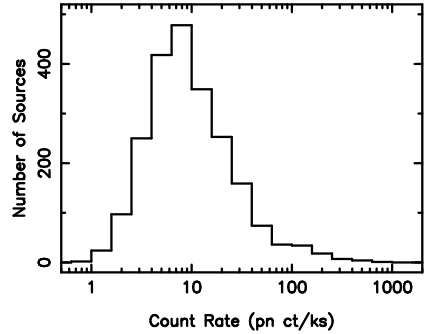}}}
\caption{The distribution of the total (soft$+$hard) count 
rate for the source sample in units of pn ct ks$^{-1}$ (0.5-12 keV).}
\label{fig:countrates}
\end{figure}


\begin{figure}
\centering
\rotatebox{-90}{\scalebox{0.50}{\includegraphics{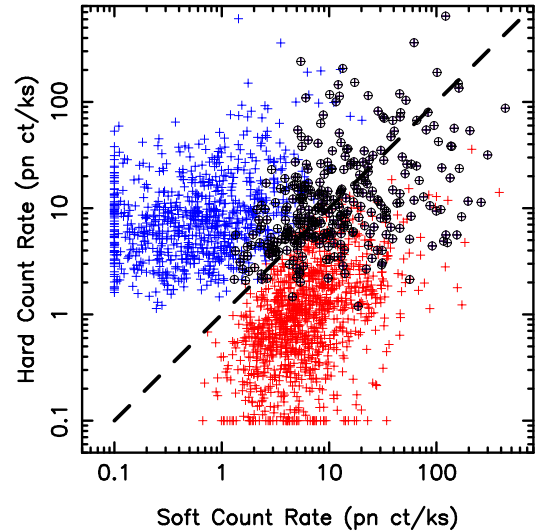}}}
\caption{Soft-band count rate versus hard-band count rate for the 
source sample. Soft-band only detections are shown as red crosses and
hard-band only detections as blue crosses. Sources detected in both 
the soft and hard bands are shown as black circles. The thick dashed
line corresponds to a hardness ratio $\rm HR=0$ (see text). Count 
rates below 0.1 pn ct/ks are plotted at this value.}
\label{fig:hrcount}
\end{figure}


\subsection{X-ray Spectral properties}
\label{sec:spec}

Fig. \ref{fig:hrdist} shows the distribution of the broad-band hardness ratio, 
HR, for the source sample.  The distribution has a saddle-like form, with 
a broad minimum centred around HR$=$0 separating maxima at, or near, 
the two extremes. We note that our focus on  source detections in 
either the soft or hard bands (or both) might introduce a bias against
sources of intermediate hardness in which the count rate is spread
fairly evenly across the two spectral channels. However, when we
plot the distribution in Fig. \ref{fig:hrdist} including only 
relatively bright sources (\eg sources with a total count rate greater than 
20 ct/ks) the saddle distribution remains broadly unchanged. We conclude that
the underlying source population splits rather cleanly into the designated soft
and hard categories.
   

\begin{figure}
\centering
\rotatebox{-90}{\scalebox{0.45}{\includegraphics{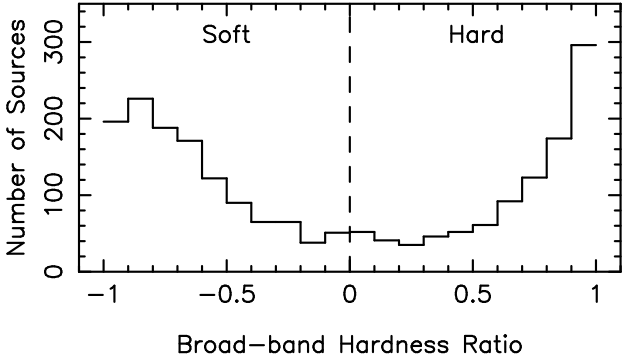}}}
\caption{Distribution of the sources with respect to
the broad-band hardness ratio, $\rm HR$.}
\label{fig:hrdist}
\end{figure}


We have also investigated the spectral distribution of the sources
with respect to the four 2XMMi bands, namely Bands 2-5.  In this context a
standard approach is to use X-ray two-colour diagrams, where each axis 
represents a different hardness ratio calculated for a pair of (adjacent)
bands. However, for our sample which is strongly affected by absorption, the
tendency was for such diagrams to be either dominated by errors (due
to low intrinsic count rates in both selected bands) or, at least, 
very compressed along one axis. We have avoided this problem by calculating 
instead a set of {\it Band Index} ($BI$) values, where $BI$-N is calculated 
simply as the count rate in Band N normalised to the total source 
count rate (\ie in this case the 
sum of the count rates in Bands 2--5). The comparison of {\it BI} 
values for adjacent bands - see Fig. \ref{fig:bandindex},  
then provides very similar information to a set of two-colour hardness 
ratio diagrams.  We note that ``quantile'' analysis, as described by \citet{Hong04},
represents an alternative strategy for investigating the spectral properties
of source samples in the low count rate limit.

Each $BI$ diagram  in Fig. \ref{fig:bandindex} shows a smoothly
varying distribution representing the range of source characteristics 
from the hardest through to the softest spectral forms. A rough calibration
of this spectral diversity is provided by the model curves 
in Fig. \ref{fig:bandindex}, which correspond
to power-law spectra with $\Gamma$ ranging from
0.5--3, subject to absorption ranging from $N_H = 
10^{20-24}~\rm cm^{-2}$. However, the sources with the softest 
spectra are not quite accommodated by the limits $\Gamma=3$ and 
$N_H=10^{20}~\rm cm^{-2}$ in the $BI$-2 versus 
$BI$-3 diagram and are, in fact, better represented 
by a thermal (Mekal) spectrum with the same limiting $N_H$ and 
$\rm kT \approx 0.5$ keV (the dashed curve in the 
top panel of Fig. \ref{fig:bandindex}). From the $BI$-3 versus $BI$-4 diagram, 
it is evident that transition from our soft to hard characterisation
occurs for $N_H = 3\times10^{21} - 3\times10^{22}~\rm cm^{-2}$ (depending
on the slope of the continuum). In the hard spectral
regime represented by the $BI$-4 versus $BI$-5 diagram, one finds that
the bulk of the signal is contained in Bands 4 and 5 for $N_H > 
3 \times 10^{22}~\rm cm^{-2}$. Evidently, there are a significant 
number of sources with $BI$-5 $>0.7$, implying $N_H$ in 
excess of $10^{23}~\rm cm^{-2}$,  provided the underlying
continuum is not exceedingly flat ($\Gamma~_{\sim}^{<}~0.5$).


\begin{figure}
\centering
\includegraphics[width=6.5cm,angle=0]{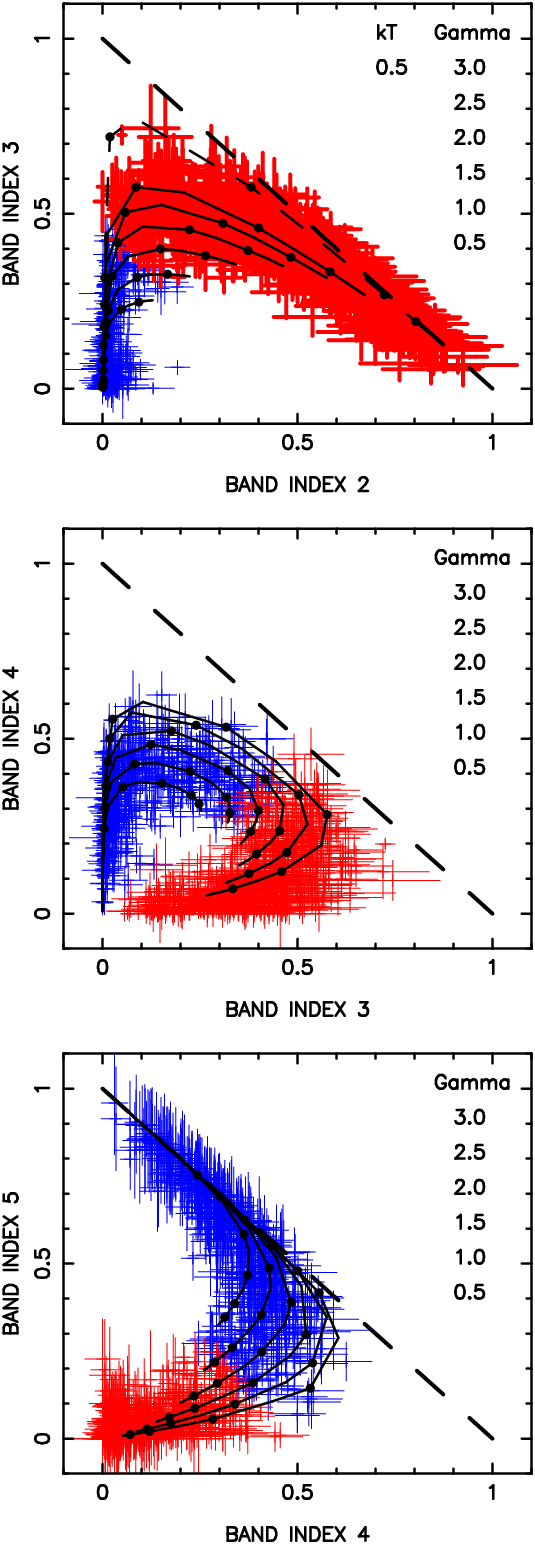}
\caption{X-ray two colour diagrams in the form of {\it Band Index} ($BI$)
plots. Here {\it Band Index} N is simply the count rate in band N
divided by the total count rate summed across the 2XMMi Bands 2--5. 
Sources categorised as soft and hard 
are shown as red and blue respectively. For clarity only sources 
with a {\it signal-to-noise} threshold in excess of six are plotted.
The solid curves represent the locus for a given photon index, $\Gamma$,
as the column density $N_H$ varies from  
$10^{20}~\rm cm^{-2}$ to $10^{24}~\rm cm^{-2}$ (with the dots
marking in an anti-clockwise order:  $10^{21}$; $3\times 10^{21}$; $10^{22}$;
$3\times 10^{22}$ and $10^{23}~\rm cm^{-2}$). The values of
$\Gamma$ are as indicated (ordered such that the higher $\Gamma$ 
loci are at a further distance from the origin).
The top panel also shows the locus (dashed curve) for a thermal Mekal
component with $\rm kT = 0.5$ keV, with the same $N_H$ 
range as above. 
The dashed diagonal lines represent an upper bound at which the full 
count rate is accounted for by the two bands in question.}
\label{fig:bandindex} 
\end{figure}


\subsection{Column density constraints}
\label{sec:nh}

We have used the measured $BI$ values to set more
specific constraints on the line-of-sight column density $N_H$
for each source.  Assuming an absorbed power-law spectral model,
we calculated a two-dimensional array of $BI$
values (one such array for each of the four spectral bands 2--5),
with log $N_H$ along the x-axis 
(log $N_H = 20 - 24 ~\rm cm^{-2}$ in steps of 0.1) and the assumed 
$\Gamma$ along the y-axis ($\Gamma = 0.5-2.5$ in steps of 0.1).
For a given source, we identified the region in each array
bounded by the {\it measured} $BI$ value and its error
bar (we actually used $1.65\sigma$ errors corresponding roughly to 90\% 
confidence combined in quadrature with a 2\% systematic).
The full constraint was then derived by  taking the minimum bound 
on log $N_H$ consistent with all four $BI$ plots. 
We set the final estimate for log $N_H$ as the mid-point
of the bound and took the half-width of the bound as the error.  
This simple recipe worked well\footnote{The procedure is, of course, only
an approximation to detailed model fitting of the available
X-ray spectral data. Nevertheless it provides very useful
indicative results.} when applied to the hard 
source sample, but was of more limited value for
soft sources, where the range of power-law models represents
a poor match to the underlying (thermal) spectra. 

The result of applying the above procedure to the hard
sample is illustrated in Fig. \ref{fig:xray_nh}(a),
which shows the X-ray derived $N_H$ values versus the source
broad-band hardness ratio, HR. 
Although the errors for individual sources are typically quite large
($\pm0.45$ in log $N_H$, on average), there is a clear trend with
$N_H$ ranging from $\sim 10^{21}~\rm cm^{-2}$ to 
$ >10^{23}~\rm cm^{-2}$ as HR varies from 0--1.
Fig. \ref{fig:xray_nh}(b) shows the comparison of the 
column density derived from the X-ray measurements with the
total Galactic column density in the source direction,
as inferred from the infrared dust measurements  of COBE/DIRBE and
IRAS/ISSA (\citealt{Schlegel98}).
From this plot it is evident that, for hardness ratio values 
HR $^{>}_{\sim}$ 0.8, which represents 50\% of the hard sample, 
the X-ray $N_H$ is very comparable to the 
total Galactic column density. In this regime, the X-ray source could be
either embedded in a localised high density region (perhaps intrinsic
to the X-ray source), located at a significant distance across
the Galactic Plane or be an extragalactic interloper. 
We consider the latter possibility in section \ref{sec:hard_pop}.  

Finally, we note that the X-ray $N_H$ values derived for the hard
source sample show a clear dependence on Galactic longitude and
latitude, albeit with a large point-to-point scatter. Sources located
within 2.5\deg of the Galactic Centre have $N_H$ values typically
a factor 2 higher than those found outside of this region, the average in
the Galactic Centre direction being $N_H = 10^{22.8}~\rm cm^{-2}$. 
Similarly, sources located within 1\deg of the Galactic plane
typically have four times the column density of sources which are
1\deg -- 2.5\deg off the plane.


\begin{figure}
\centering
\includegraphics[width=7.5cm,angle=-90]{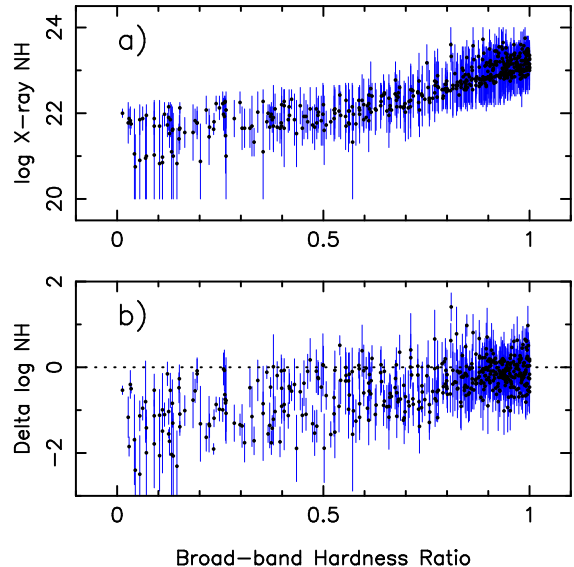}
\caption{a) The line-of-sight $N_H$
estimated from the $BI$ values plotted 
against the broad-band hardness ratio, HR. Only sources categorised as 
hard and with a {\it signal-to-noise} threshold $>$ 6
are shown. An underlying power-law continuum 
spectrum, with $\Gamma$ in the range 0.5--2.5, is assumed. 
b)  A comparison of the X-ray derived $N_H$  
with the total Galactic $N_H$ in the source direction 
determined from infrared dust measurements. The y-axis represents the difference
between  the X-ray and the infrared estimates of log $N_H$.}
\label{fig:xray_nh}
\end{figure}


\begin{figure}
\centering
\includegraphics[width=5.5cm,angle=-90]{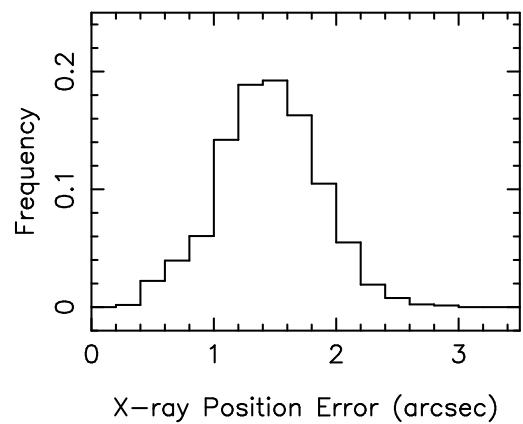}
\caption{The distribution of X-ray ($1\sigma$) position errors
for the source sample.}
\label{fig:xerrors}
\end{figure}


\section{Near-Infrared Counterparts}

The identification and characterisation of longer-wavelength counterparts 
represents a crucial step in determining the nature of Galactic X-ray sources. 
The ideal starting point for this would be a sub-arcsecond imaging survey in 
a set of wavebands not too strongly influenced by Galactic absorption, with 
commensurate high precision astrometry.  At present a comprehensive 
{\it dedicated} image database of this quality is not available for the 
{\sl XMM-Newton} survey fields. However, we have conducted a systematic 
investigation  based on cross-correlations with two near-infrared surveys which are 
available, namely 2MASS and UKIDSS.

\subsection{Cross-correlation with 2MASS}
\label{sec:2mass}

The Two Micron All Sky Survey (2MASS)  (\citealt{Cutri03}; \citealt{Skrutskie06}) 
provides uniform 
coverage of the entire sky in three near-infrared (NIR) bands, namely in 
J (1.25$\mu$), H (1.65$\mu$) and K$_{s}$ (2.17$\mu$) on a scale of 
2.0\arcsec (this is the pixel size of 2MASS images; the FWHM of the PSF was
typically 2.5\arcsec-3.0\arcsec). At~~high latitude the nominal 
survey completeness is 15.8, 15.1 
and 14.3, respectively. However, in the Galactic Plane the measured source 
counts turn down at limits one magnitude (or more) brighter, because of the 
effects of source confusion on the detection thresholds. The astrometric 
uncertainty is generally less than $0.2\arcsec$, 
although this may be compromised in very confused regions.
 
As a first step, we cross-correlated our X-ray source sample with the 2MASS 
catalogue and, for each X-ray source position, extracted the set of 2MASS 
sources within a radius of 20\arcsec. A total of 27485 2MASS sources 
were associated with the 2204 X-ray source positions via this 
process.
Here we use the X-ray source position and position error quoted 
in the 2XMMi catalogue for the individual source observations (parameters: 
{\it RA, DEC, poserr}), where the position error includes a systematic
error added in quadrature with the statistical error (the former
was set to 0.35\arcsec~if the field astrometry was corrected by reference 
to the USNO B1.0 optical catalogue, or 1\arcsec otherwise).
Fig.\ref{fig:xerrors} shows the distribution of X-ray position
errors for the source sample; the range is from 0.37\arcsec 
to 3\arcsec, with an area-weighted average of 1.5\arcsec. In the event
we selected only those X-ray sources with X-ray position errors
$<$2\arcsec for the cross-correlation study, resulting in a
reduced sample size of 2016 X-ray sources. 

A preliminary investigation of the incidence of 2MASS sources
at or near the X-ray positions demonstrated that, at least for 
the soft X-ray source sample, there was a significant excess 
over the number expected by chance, presumably reflecting the presence 
of a real counterpart in a substantial number of cases\footnote{In 
establishing the coincidence rate of 2MASS sources with the soft and hard
X-ray source samples, we did not employ any discrimination based on the 2MASS
quality flags. However, the flags were use to select sources with good 
photometry in the later investigation of the impact of interstellar 
reddening on the stellar colours.}. 
The distribution of this excess population with respect to the X-ray 
source positions was studied by considering how the net number of
sources contained within the soft-source error circles (over and above 
the background rate) varied as a function of the assumed X-ray error 
circle radius (Fig.\ref{fig:cumul}).
On the basis of this empirical information, we set the X-ray error 
circle radius to be 2.5 times the X-ray position error 
quoted in the 2XMMi
catalogue, at which point  90\% of potential {\it real} 
counterparts will be contained within the error circle.


\begin{figure}
\centering
\includegraphics[width=6cm,angle=-90]{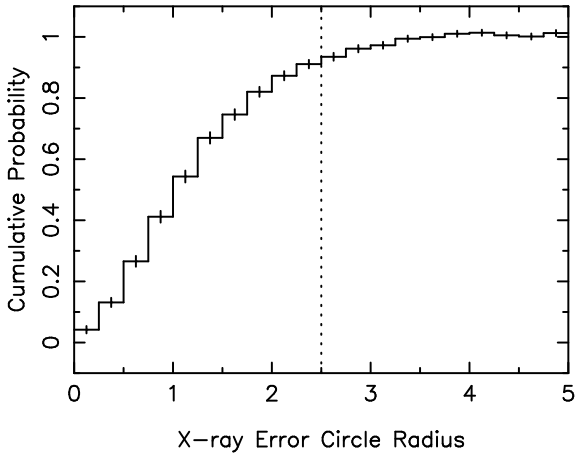}
\caption{The cumulative probability of finding an associated 2MASS
source in an X-ray error circle, plotted as a function of the 
error circle radius (in units of the $1\sigma$ X-ray position error). 
The analysis is restricted to 2MASS sources with J $= 5-15$ found 
in the fields of the soft X-ray sources. 
For the current work, the radius of the X-ray error circle was 
taken to be 2.5 times the $1\sigma$ X-ray position error 
(as indicated by the vertical dotted line), thereby encompassing 
around 90\% of the likely NIR counterparts.}
\label{fig:cumul}
\end{figure}


For the soft X-ray sources, the number of error circles containing a
2MASS source was 987 out of 1120 ($88\%$), whereas for 
the spectrally hard sources the statistics were 393 out of 896
($44\%$). In most instances, there was only one 2MASS source 
within the X-ray error circle, but for the 85 X-ray sources for which 
this was not the case, we treated the brightest 2MASS source 
(in J band) as the potential counterpart.  Figure \ref{fig:id_frac} 
shows how the incidence rate of 2MASS stars within the  X-ray error 
circles accumulates as a function of the 2MASS J magnitude for
both the soft and hard source categories. 
The chance coincidence rate, calculated using the 2MASS catalogue information  
for the field regions within 8--20\arcsec of the X-ray 
source position, is also indicated.  All three curves begin to flatten and 
eventually reach a plateau, around J $\approx 16$, consistent with the fact 
that the spatial density of sources per unit magnitude in this set of 
Galactic fields peaks at J $\approx 15$ (\ie at a level roughly a magnitude 
brighter than the completeness limit of the 2MASS survey at high latitude, 
as noted earlier). The plateau level for soft and hard distributions is
as quoted above, whereas for the chance distribution it is $\sim 30\%$.


\begin{figure}
\centering
\includegraphics[width=5.0cm,angle=-90]{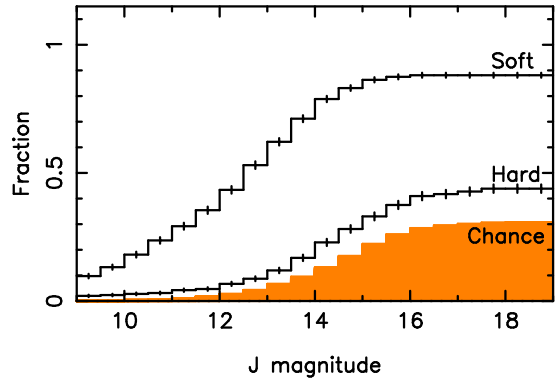}
\caption{The cumulative fraction of X-ray error circles containing a 
2MASS star as a function of the J magnitude of the star. Separate curves are 
shown for the soft and hard X-ray source samples. 
The chance coincidence rate for field stars is shown by
the filled histogram.}
\label{fig:id_frac}
\end{figure}


\begin{figure}
\centering
\includegraphics[width=7cm,angle=-90]{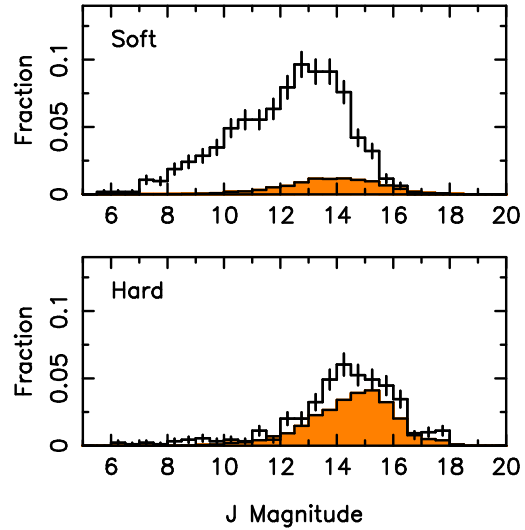}
\caption{The fraction of error circles containing 2MASS stars
as a function of the J magnitude for the soft ({\it top panel})
and hard ({\it bottom panel}) source samples. This is a differential
plot, \ie the ordinate is the number per magnitude bin
normalised to the total number of sources in the 
soft/hard sample. The predicted distribution of chance coincidences with 
field stars, based on the statistics for 2MASS stars located 
at radial offsets between 8\arcsec--20\arcsec~~ 
from the X-ray position, is shown as the filled histogram.  This prediction
takes into account the declining fraction of error circles which are available 
at faint magnitudes due to the presence of brighter stars.}
\label{fig:magn_dist}
\end{figure}


Figure \ref{fig:magn_dist} shows the derived magnitude distribution for 
the {\it brightest} 2MASS stars found to be positionally coincident 
with the X-ray source positions. The predicted contamination due to chance 
coincidences is also shown,  where a correction has been incorporated for 
the fact that an error circle may not be ``available'' at faint magnitudes
if a brighter star is already present.
For the soft X-ray source sample, the magnitude distribution extends from
J=7-16.  Down to J=16,
the underlying contamination due to field stars amounts to roughly 9\% 
out of the 88\% of error-circles which contain 2MASS stars down to
this magnitude limit.
In contrast, the magnitude distribution for 
the hard sample shows a much more modest excess of coincidences
with respect to the predicted chance rate; 
for J~$<$~16, the coincidence rate is 38\% of which
24\%, \ie roughly two-thirds, can be attributed to contamination by 
field stars. 

The probability of finding a positionally coincident 2MASS source is
clearly very different for the soft and hard 
X-ray source samples.  We have investigated this in more detail by 
considering how this probability varies as a function of the 
broad-band hardness ratio, HR.  Fig. \ref{fig:hr_ids} shows the 
result.  The probability of finding a 2MASS star brighter than $J=16$
in the X-ray error circle (inclusive of chance hits) exhibits
a step function behaviour, remaining at close to 90\% for the spectrally 
soft sources with HR in the range -1 to -0.1, followed by a rapid 
transition to a level of about 50\% for HR values near 0 
followed by further decline to a level very near to the chance rate 
for the sources with the hardest spectra.


\begin{figure}
\centering
\rotatebox{-90}{\scalebox{0.45}{\includegraphics{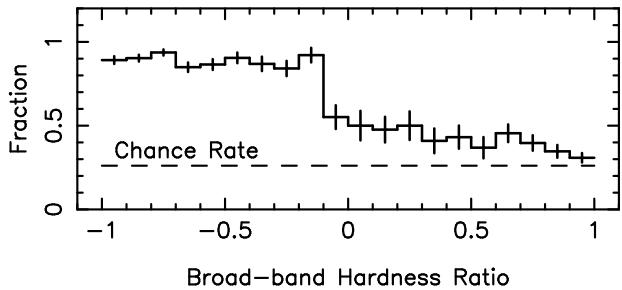}}}
\caption{The fraction of sources with a 2MASS source 
with J$<16$ in the error-circle as a function of $\rm HR$. 
The chance coincidence rate is also indicated.}
\label{fig:hr_ids}
\end{figure}



\subsection{Cross-Correlation with UKIDSS}
\label{sec:ukidss}

For our application, the 2MASS survey becomes incomplete below
J $\approx 15$.  In contrast the United Kingdom Infrared 
Deep Sky Survey (UKIDSS, \citealt{Lawrence07}) provides better spatial 
resolution (0.2\arcsec pixel scale with typically $\sim$ 1\arcsec seeing)
and is significantly deeper. The nominal
limiting magnitudes estimated for the UKIDSS survey are
J=19.77, H=19.00 and K=18.05, with uncertainties of up
to $\sim$ 0.2 magnitudes. As with 2MASS, the appropriate
limits for the Galactic plane will be considerably brighter
depending on the actual crowding within a given field
(\citealt{Lucas08}). The photometric 
data used in this analysis is taken from the 
UKIDSS Galactic Plane Survey (GPS) Data Release 4.

UKIDSS GPS J-band data were available for a total of 1117 X-ray sources
drawn from our sample of X-ray sources with position errors
less than 2\arcsec, with the coverage split roughly equally 
between the soft and hard samples (\ie 596 soft and 577 hard). For this
analysis we extracted a preliminary set of UKIDSS stars within
a nominal $10\arcsec$ of each X-ray position, leading to a preliminary set
of 36974 stars down to J=20. Using the same X-ray error circle radii 
as for the 2MASS study, we repeated the analysis of the coincidence rates
of UKIDSS stars within the X-ray error circles. As before, we make the
assumption that the brightest star in the circle (in J band) is the most likely 
counterpart.

As we push down to a J magnitude of 20, the incidence rate
of infrared stars in the X-ray error circles rises to
97\% for the soft sources and 92\% for the hard sources.
Of course at these faint levels, given the stellar density and size
of the error circles,
the chance rate is commensurately high, namely 85\%. 
Figure \ref{fig:magn_dist_uk} shows the resulting magnitude distributions 
for the {\it brightest} UKIDSS star in the X-ray error circle. 
 
For the soft sources, the magnitude distribution is 
truncated below J $< 10$ as a result of saturation effects 
in the UKIDSS images, with the apparent sub-peak near $J \approx 11$
presumably attributable to the same effect.  As was evident in the 
equivalent 2MASS distribution, there is a sharp cutoff between J=15-16. 
Despite the fact that the UKIDSS survey goes between 
3-4 magnitudes deeper than 2MASS, the evidence is that
very few {\it real} soft-source counterparts have J $> 15.5$. 
For the soft sources, the coincidence rate with
bright UKIDSS stars (J $<16$) is 91\% inclusive of a chance
rate of 9\%, values which are fully consistent with the
equivalent 2MASS estimates. 

For the hard sources, the coincidence rate with
bright UKIDSS stars (J $<16$) is 45\%, which splits into a
likely chance rate of 29\% and an excess rate (\ie
potential real counterparts) of 16\%.  These percentages are again very 
comparable to those obtained from the 2MASS cross-correlation. 
There is also evidence that real counterparts are seen 
at fainter magnitudes, with the J=16-20 coincidence 
rate of 47\% comprising a likely chance rate of 38\% 
and an excess rate of 9\%. It follows that $\sim  25\%$ of the
hard source population may have counterparts visible, as the 
brightest star in the error circle, down to J=20. Of course,
for every error circle for which this is the case, there are
roughly 2.7 instances where the brightest star is just a chance
positional coincidence of a field star. Clearly this
statistic implies detailed follow-up of the hard-source 
population carries a large overhead. 


\begin{figure}
\centering
\includegraphics[width=7cm,angle=-90]{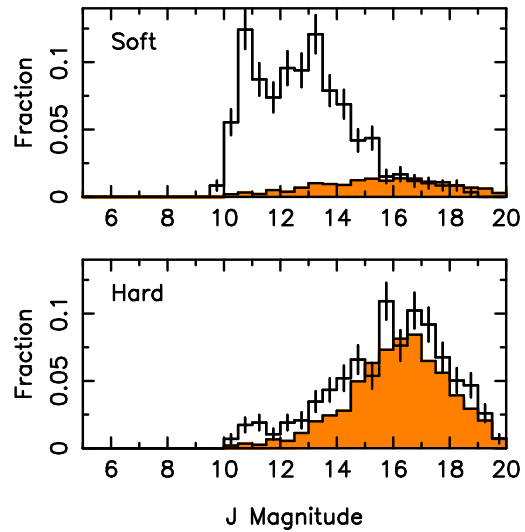}
\caption{The fraction of error circles containing UKIDSS stars
as a function of the J magnitude for the soft ({\it top panel})
and hard ({\it bottom panel}) source samples.  This is a differential
plot, \ie the ordinate is the number per magnitude bin, normalised 
to the total number of sources in the soft/hard sample.
The predicted distribution of chance coincidences with field stars, based on
the statistics for UKIDSS stars located at radial offsets 
between 7.5\arcsec--10\arcsec~~ from the X-ray position, is shown as the 
filled histogram.  This prediction takes into account the declining fraction 
of error circles which are available at faint magnitudes due to the presence 
of brighter stars.}
\label{fig:magn_dist_uk}
\end{figure}


\subsection{Reddening of potential stellar counterparts}
\label{sec:colours}

In order to investigated the colours of the NIR stars found 
inside the X-ray error circles, we have carried out a
further selection aimed at removing those objects 
affected by known issues in relation to their photometry.
The 2MASS catalogue includes four quality flags. For our purpose, 
we utilise the so-called the photometric {\it QUALITY} flag  and select only
those stars flagged as {\it AAA} (\ie valid measurements with signal-to-noise ratio $>$10
in J, H and K).
Similarly for the UKIDSS stars we required the error bit flag,
{\it ppErrbits} $<$ 256,  for each of the three near-IR bands
and also $pstar \ge 0.99$ (\ie the removal of sources with
non-stellar profiles).  A total of 840 2MASS stars, representing 
the brightest object in the X-ray error
circle, passed this selection step, with the equivalent number for UKIDSS
being 444. A total of 130 stars were common to both lists and, in the event,
we used only the 2MASS data for this subset of sources.


\begin{figure}
\centering
\includegraphics[width=7cm,angle=-90]{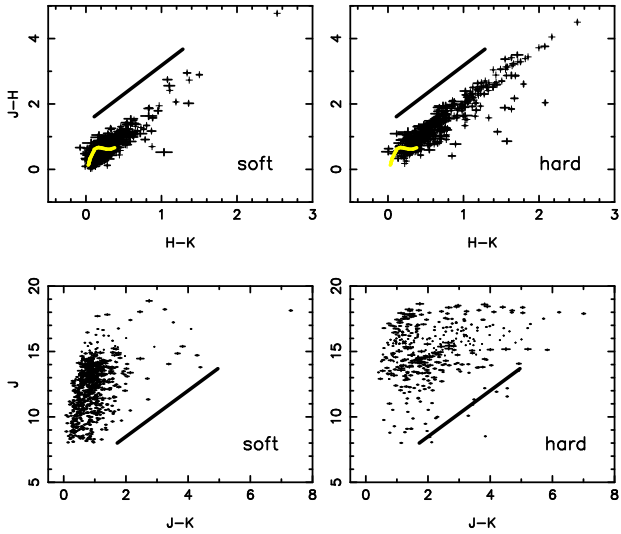}
\caption{{\it Top panels:} The H-K versus J-H colours of the brightest stars
in the X-ray errors circles, shown separately for the soft and hard X-ray source
samples.  The curved line is the locus of  
dwarf stars from F0V to M6V. The diagonal line shows the reddening vector 
for $A_V = 20$.
{\it Bottom panels:} The J-K colour versus J magnitude 
of the brightest stars in the X-ray errors circles,  shown separately for the soft 
and hard X-ray source samples.  
The diagonal line shows the impact of $A_V = 20$ on the J-K colour 
and J magnitude of a given star.
}
\label{fig:colours_all}
\end{figure}


\begin{figure}
\centering
\includegraphics[width=5cm,angle=-90]{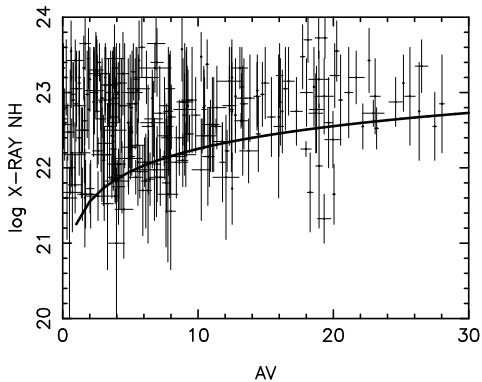}
\caption{A comparison of the visual absorption $A_V$ determined 
from the 2MASS/UKIDSS stellar colours with the estimated line-of-sight
gas column density $N_H$ derived from the X-ray spectral characteristics.
The curve shows the standard relationship $N_H = 1.79 \times 10^{21}$ $A_V$
$\rm ~cm^{-2}$ (\citealt{Predehl95}).}
\label{fig:av_nh}
\end{figure}


Figure \ref{fig:colours_all} shows the H-K versus J-H two-colour diagram
for the combined 2MASS/UKIDSS\footnote{We have not applied the photometric
transformations detailed in \citet{Hodgkin09} to convert the UKIDSS measurements
to the 2MASS photometric scale, since the scatter in the plots far exceeds 
the magnitude of the corrections.} sample of coincident stars, with the results
for the soft and hard X-ray sources shown separately. The locus of main-sequence 
dwarf stars from F0V to M6V (\citealt{Pickles98}) and the reddening vector
for a visual absorption $A_V = 20$ are also defined. For the stars potentially
associated with soft X-ray sources, there is a dense grouping in the
region of the two-colour diagram encompassed by the locus of
non-reddened main sequence dwarf stars, with a spread along the
reddening direction largely encompassed by an $A_V$
of up to 20. In the case of the hard X-ray sources, the implied
reddening is much more substantial for many of the stellar candidates,
although the bulk of the population are still bounded by $A_V <20$. 
Particularly for the hard band sample, there are a number of stars which 
lie off to the right of the main distribution (\ie have very red H-K colours);
these may be pre-main sequence (T Tauri) stars exhibiting an infrared excess
(\cf \citealt{Koenig08}).

Figure \ref{fig:colours_all} also shows the variation of the J magnitude versus
the J-K colour, with the impact of a visual absorption $A_V = 20$ 
again indicated. The absolute magnitude, $M_J$,  
for main sequence stars of type F0V to M0V  ranges from 2.43--5.72
(\citealt{Covey07}), \ie a $\Delta J \approx 3.3$ for unreddened stars 
at a fixed distance. If we compare this with the spread of $\Delta J\approx7$ 
apparent in the colour-magnitude diagram for the bulk of the stars 
associated with the soft X-ray sources, 
then the implied scatter in distance is roughly a factor of 5. Of course, the 
presence of significant numbers of dMe stars of spectral class later 
than M0 would compromise this argument. The stars found in the hard X-ray source
error circles are significantly fainter and redder than the population
associated with the soft X-ray sources.  However, as noted earlier, 
a substantial fraction of the stars linked to the hard X-ray sources
will be chance coincidences with field stars and to first order
the scatter in the both the H-K versus J-H two-colour diagram
and in the J-K versus J colour-magnitude diagram for the hard 
sources must reflect the underlying properties of the field star distribution.
 
For those stars linked to hard X-ray sources, we have estimated the 
visual absorption by projecting the star's position in the two colour diagram 
onto the reddening vector and determining the offset relative to an origin at 
J-H=0.61, H-K=0.11 (the colours of an unreddened GV star). 
In Fig. \ref{fig:av_nh}, the resulting estimate of $A_V$ is plotted versus the X-ray 
column density derived in \S4.4.      Compared to the standard conversion,
$N_H = 1.79 \times 10^{21}~A_V$ (\citealt{Predehl95}), most of the points in the diagram lie
at $N_H$ values
significantly greater than that implied by the $A_V$ determination. 
This is to be expected if many of the associated NIR objects are, 
in fact, foreground stars, 
albeit sufficiently distant to be subject to significant reddening.  
One way of picking out potentially {\it true} counterparts 
might be to require the derived $A_V$ and $N_H$ estimates to be comparable;
unfortunately the large uncertainties implicit in the estimation of both
parameters mitigates against this as a practical scheme. 

\vspace{4mm}

\section{Discussion}

\subsection{Nature of the Soft X-ray Source Population}
\label{sec:soft_pop}

In the soft band, the  nominal sensitivity limit of our survey is
around 2 pn ct/ks corresponding to an {\it unabsorbed} flux,
$F_X = 4 \times 10^{-15}~\rm erg~cm^{-2}~s^{-1}$ (0.5-2 keV).  
This is roughly 5 times deeper than 
the ROSAT survey reported by \citet{Morley01}, which in turn was a factor
5 deeper than the {\it Einstein} GPS (\citealt{Hertz84}; 
\citealt{Hertz88}) and the {\it ROSAT} GPS (\citealt{Motch91}; \citealt{Motch97}).
By way of comparison, the stellar survey carried out recently under the auspices 
of the {\it Extended Chandra Multiwavelength Project} (\citealt{Covey08})
reaches similar, or somewhat deeper, limits to those reported here.
These surveys and many other studies have established
that the Galactic source population in the soft X-ray band 
largely comprises of late-type stars with active coronae.
More specifically the ROSAT all-sky survey clearly demonstrated
that main-sequence stars from early-F to late-M dominate the source 
statistics and that stars younger than $\sim2$ Gyr are typically
more luminous in X-rays than older populations; this is a consequence of the 
reducing efficiency with which the corona is heated by the dynamo mechanism as 
the rate of stellar rotation declines with age (see \citealt{Gudel04} and references therein). 
At high galactic latitude, as the flux limit is lowered, the
surveyed volume will eventually extend beyond the scale height of the young 
star population, thereby altering the balance between relatively old and 
relatively young stars (\eg \citealt{Micela07}). Although this is not a 
consideration which applies directly to a narrow Galactic plane survey, 
line-of-sight absorption will suppress the surveyed volume most severely
for higher $L_{X}$ sources, presumably giving rise to a similar bias. 

From the cross-correlation of our soft X-ray source sample with 
the 2MASS catalogue, we concluded that 88\% of the error circles 
contain stars with J $<$ 16, with only 9\% of this figure
attributable to chance coincidences. Furthermore, since our error 
circles were scaled so as to contain $\sim$90\% of the {\it true}
counterparts, it follows that active stellar coronae 
must account for almost our entire sample of soft X-ray 
sources. When the search for counterparts was extended 
some 4 magnitudes fainter through the 
cross-correlation with the UKIDSS catalogue, the  evidence for a cut-off 
at J $\approx$16 was quite stark\footnote{If we compare the
corresponding X-ray and infrared flux limits, 
$F_X = 4 \times 10^{-15}~\rm erg~cm^{-2}~s^{-1}$ and J = 16,
through the relation log$(F_{X}/F_{J}) = $log$(F_{X})$ + 0.4~J + 6.30, 
we obtain log$(F_{X}/F_{J}) = -1.7$.  We note that ROSAT sources correlating with 
stellar counterparts extracted from either the Simbad or Sloan
databases also show a cutoff in their X-ray to infrared flux ratio
at or near this limit (\citealt{Agueros09}).}. This prompts the question
of why the stellar identifications are restricted to bright magnitudes?

In Fig. \ref{fig:dist_limits} we explore the distances out to
which main sequence stars of different spectral class may be detected in 
both the NIR and our {\it XMM-Newton} survey. In the NIR, the constraint
J $\le 16$ sets a distance limit which varies from 
2.7 kpc for early F stars down to $\sim 70$ pc
for late M stars.  In contrast if we assume stellar X-ray luminosities
typical of normal solar-type stars, as represented
by the NEXXUS survey (\citealt{Schmitt04}), then the X-ray survey
horizon is an order of magnitude closer. Of course, 
as the level of coronal activity increases, so does the 
surveyed volume in the X-ray band. As is clear from
Fig. \ref{fig:dist_limits},  if the X-ray horizon is to match 
the J=16 boundary, the X-ray luminosity will need to 
have increased to the point at which
stellar-coronal emission is known to saturate, namely at
$L_X = 10^{-3} \times L_{bol}$ (\citealt{Gudel04} and references 
therein).  This, almost certainly, explains the origin of J $\sim$ 
16 cut-off alluded to earlier.  

It is true that previous X-ray surveys have identified 
some coronal sources emitting above the nominal saturation limit, 
one explanation being that these are systems caught whilst flaring  
(\eg \citealt{Fleming95}; \citealt{Morley01}). \citet{Fleming95} 
estimate that $\sim$25\% of solar-like stars and $\sim$50\% of dMe stars were 
detected in the Einstein GPS, whilst flaring, so such sources could 
represent a not insignificant fraction of the population in our current sample.
However, on the basis of the above evidence it would seem that
even allowing for stellar flaring, the saturation limit is
applicable to the vast majority of our soft band detections. Similar
arguments and constraints  apply in the case of tidally interacting close binaries, such as RSCVn systems,
which maintain fast rotation throughout their main-sequence lifetime possibly into
their later evolution (\citealt{Gudel04}) and which are amongst the
most luminous coronal emitters, with $L_X$ in some cases exceeding $10^{31} \rm~erg~s^{-1}$ 
(\eg \citealt{Makarov03}).

Figure \ref{fig:dist_limits} also shows an additional X-ray constraint at about
3 kpc, which is the distance at which the line-of-sight column density exceeds 
$N_H \sim 1.2 \times 10^{22} \rm~cm^{-2}$ (albeit based on a very simplistic
absorption model as detailed in the figure caption). At this value of $N_H$ the soft X-ray 
absorption is such as to move a source from the soft to the hard category
(assuming a powerlaw spectrum with $\Gamma=2.5$). The implication
is that there is a ``spectral'' horizon for the soft X-ray source sample 
at about 3 kpc, which corresponds to a soft-band  X-ray luminosity of
somewhat less than $10^{32} \rm~erg~s^{-1}$.  On the basis of the saturation 
limit for coronal emission, it would seem that, at least amongst single stars,
only F stars are likely to approach the $N_H$ horizon. Conversely, one might conjecture
that relatively distant F stars are likely to be predominant as one approaches 
the HR=0 boundary.  

Finally, we note that the soft source population could also include
some accretion-powered stellar systems, in particular CVs containing
only weakly magnetized white dwarfs. Such sources comprise $\sim 90\%$
of the accreting white dwarf population (\eg \citealt{Warner95}),
have relatively soft thermal components in their spectra and soft X-ray luminosities 
typically in the range $L_X \sim 10^{29-32} \rm~erg~s^{-1}$
(\eg \citealt{Verbunt97}). Presumably non-magnetic CVs might be 
detected out to the $N_H$ horizon and given the low-mass nature
of the secondaries be rather faint in the near-infrared. However, 
given the above statistics, this class of object can, at best, 
represent only a small fraction of the soft source population. 


\begin{figure}
\centering
\includegraphics[width=6.5cm,angle=-90]{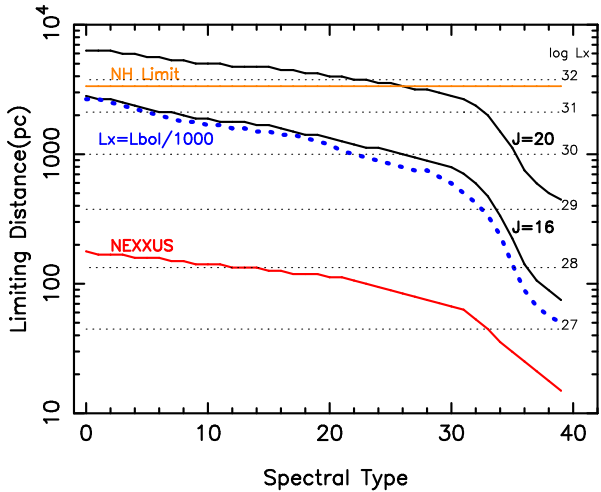}
\caption{The limiting distance out to which sources can be detected
plotted as a function of the stellar-spectral class. Here spectral
type 0--39 corresponds to dwarf stars from F0V--M9V, with 
interpolation across spectral classifications, as necessary. 
The thick black lines correspond to the distances at which stars
reach limiting J magnitudes of 16 and 20.  The visual absorption,  $A_V$,
in the Galactic plane was assumed to increase at the rate 2 mag/kpc,
comparable to that inferred from the UKIDSS GPS survey data
for ($l,b$)=(31,0) (see \citealt {Lucas08}),
with the conversion $A_J=0.286 \times A_V$ (\citealt{Fiorucci03}). The horizontal grid 
illustrates the distances for a given X-ray luminosity out to which
sources may be detected assuming an
effective X-ray survey limit of $F_X = 4 \times 10^{-15}~\rm
erg~cm^{-2}~s^{-1}$ (0.5-2 keV). 
The soft X-ray absorption in the Galactic plane was determined assuming
$N_H = 1.79 \times 10^{21}$ $A_V \rm ~cm^{-2}$ and a power-law
source spectrum with $\Gamma=2.5$. The lower (red) curve
shows the distance to which sources may be detected
with the median $L_X$ of nearby stars as represented by the 
NEXXUS catalogue (\citealt{Schmitt04}). The dotted (blue) curve 
represents the distance out to which stars might be detected
assuming $L_X = 10^{-3} \times L_{bol}$.  The horizontal line
labelled ``NH limit'' represents the distance at which, due
to absorption, sources switch from a soft to a hard characterisation.
Absolute $M_J$ and $M_{bol}$ values for dwarfs stars were taken 
from Table 5 of \citealt{Kraus07}.
}
\label{fig:dist_limits}
\end{figure}


\subsection{Nature of the Hard X-ray Source Population}
\label{sec:hard_pop}

In the hard X-ray band, if we again take the nominal sensitivity limit
to be 2 pn ct/ks, then the corresponding  {\it unabsorbed} flux limit
is  $F_X = 2.5 \times 10^{-14}~\rm erg~cm^{-2}~s^{-1}$ (2--10 keV).
This translates to a luminosity of $\rm 3 \times 10^{30}~(d/1 kpc)^{2}~erg~s^{-1}$
for a source at a distance d in kpc.  In the 2--10 keV band, our {\it XMM-Newton} 
survey is roughly an order of magnitude deeper than the ASCA 
GPS (\citealt{Sugizaki01}) but shallower by about the same factor with 
respect to the Chandra deep pointing
in the Galactic Plane reported by \citet{Ebisawa05}.

The incidence of NIR counterparts brighter than J=16 within
the X-ray error circles shows a sharp downward step around HR $=0$ 
(Fig. \ref{fig:hr_ids}), indicative of the fact that the underlying source 
populations differ markedly between the hard and soft bands.
On the basis of the 2MASS and UKIDSS cross-correlation analyses, 
it appears that {\it true} counterparts may be present, as the brightest star 
in the X-ray error circle, in roughly 25\% of cases, down to a limiting magnitude of
J=20, set against a chance rate which is 2.7 times this value.  If the cut is set 
at a brighter limit, at say J~$<$~16, then true counterparts may be present in roughly 
15\% of cases, set against a chance rate of somewhat less than twice this value,
This rather poor ``success rate'' for an X-ray survey with source locations
(or more specifically 90\% error circle radii) of better than 5\arcsec~radius is, 
of course, evidence for the relative faintness in the NIR (and optical) of the 
bulk of the true counterparts. For the hard-band sample, the average $N_H$
inferred from the X-ray spectral analysis is $\sim 3 \times$10$^{22}$ 
cm$^{-2}$, implying $A_V \sim 17$ and $A_J \sim 5$, so it is not surprising
that the NIR objects linked to the hard sources are typically very much fainter 
than the  NIR counterparts of the soft sources, for which the 
impact of line-of-sight absorption is much less severe.

In broad terms, the various classes of hard X-ray
emitting source split into systems containing either a high-mass or low-mass
star (see below). For the stellar-light component, the difference in the absolute 
magnitude will be at least $\Delta M_J \sim 5$ (\eg comparing $M_J$ for main-sequence
stars earlier than B8 with that for dwarfs later than K0; \citealt{Kraus07}),
implying a  difference in log$(F_{X}/F_{J})$ for high-mass and low-mass systems
(at a given distance emitting at the same $L_X$) of at least a factor of 2.  Even when
the total systemic emission is considered, any additional contribution to the 
light from accretion disks, hotspots, etc., generally fails to bridge this gap 
(\eg as in CVs; \citealt{Ak07}). It follows that there will be a strong bias towards
high-mass systems within the subset of sources with bright NIR 
counterparts.  Of course, for relatively high luminosity systems 
($L_X > 10^{32} \rm~erg~s^{-1}$) detectable in the hard band
out to a few kpc and beyond, the impact of interstellar absorption at J will also 
be factor tending to suppress the ``identification'' rate\footnote{Although the 
impact of interstellar absorption will be reduced at K, the commensurately higher 
stellar densities mitigates against any clear advantage in focusing on
the longer wavelength band for the current application.}.
 
Although our knowledge of the hard X-ray source population of the Galaxy at 
intermediate to faint fluxes is very incomplete, due to the difficulty in
identifying large samples of objects drawn from hard X-ray catalogues, we do know
at least in qualitative terms which classes of source may be
present in significant numbers.  The largest single contribution is likely to 
come from CVs containing an accreting magnetic white dwarf
(polars or intermediate polars) in which there is substantial
hard emission characterised by thermal temperatures of kT $_{\sim}^{>}$ 10 keV 
(\eg \citealt{Ezuka99}). This class of source which may 
account for the very high density of X-ray sources observed in the Galactic Centre  
(\eg \citealt{Muno03}) and also for a substantial fraction
of the hard unresolved X-ray emission, known as the Galactic Ridge,
observed both near the Galactic Centre
and also along the inner quadrant of the Galactic Plane 
(\eg \citealt{Sazonov06}; \citealt{Rev06}).
LMXB and HMXB systems emitting at relatively low $L_X$ may also be present 
in the survey. For example, \citet{Pfahl02} have suggested that
neutron stars accreting from the winds of main-sequence stellar 
companions might be plentiful in the Galaxy. A similar idea was suggested
by  \citet{Willems03} involving pre-low-mass X-ray binaries. However, on the basis of
a population synthesis model,  \citet{Liu06} conclude that neutron
star LMXB transients in relative quiescence, LMXB with white dwarf donors
and rotation-powered pulsars may provide an alternative explanation of the 
high density of faint X-ray sources seen in the Galactic Centre. 
Our survey which encompasses star-forming regions both in the Galactic Centre region
and  in inner spiral arms  of the Galaxy, 
may also contain new examples of the highly embedded Supergiant Fast X-ray 
Transients (SFXTs) recently discovered by Integral (\eg \citealt{Sguera06}) 
and of Be-star X-ray binaries, including $\gamma$ Cas
analogs (\eg \citealt{Motch07}).

Relatively hard X-ray emission can also be produced in stellar sources without
reliance on accretion power.  Shocks produced in the unstable winds of 
massive Wolf-Rayet (WR) and O-supergiant stars can generate emission above 
2 keV. This hard emission can be greatly enhance in WR/OB binaries, 
systems which appear to be relatively common in the Galactic Centre 
(\citealt{Mauerhan09}; {\citealt{Mauerhan10}). 
Coronal emission in close, tidally interacting, binaries such as RSCVn-type systems 
and related classes of object (BY Dra-type and Algol-type binaries)
may also reach temperatures extending into the hard X-ray regime
(\citealt{Swank81}; \citealt{Agrawal81}).
Similarly, in specific locations such as the vicinities of dense molecular clouds 
and star-forming regions, protostars and T-Tauri stars may also be prominent 
as hard X-ray sources, by virtue of their enhanced
high temperature coronal emission, particularly during flaring episodes
(\eg \citealt{Stelzer00}; \citealt{Tsu02}).
Of course, it remains to be demonstrated whether any new types of 
Galactic X-ray source are present within current hard X-ray Galactic surveys;
for example, isolated neutron stars and/or blackholes accreting directly from 
the interstellar medium (\citealt{Agol02}).

An illustration of the likely makeup of hard-band selected samples
is provided by the recent study of \cite{Motch10} who carried out an optical
follow-up campaign on a sample of sources selected from the {\it XMM-Newton}
Galactic plane survey of \cite{Hands04}. Out of 25 hard sources in the follow-up
sample,  there were tentative identifications with four HMXBs and 
and with one likely Wolf-Rayet colliding-wind binary system. 
A number of the HMXBs had particularly interesting properties, for example
the group included  a very absorbed likely hyper-luminous star 
in a system bearing similarity to $\eta$ Carina, a new/Be star binary belonging to the
growing class of $\gamma$-Cas analogs and a possible supergiant X-ray binary
of the type recently discovered by Integral. The inferred 
X-ray luminosities for these massive star systems was 
$L_X = 10^{32-34} \rm erg~s^{-1}$.  There were also probable identifications
of 3 CVs with relatively faint optical counterparts (including a previously undiscovered 
magnetic system displaying strong X-ray variability) and somewhat more tentative 
identifications of 2 LMXBs.
The detection of several potential active coronal emitters above 2 keV
also underlines the possibility that active RSCVn binaries might
contribute significantly to hard-band selected samples.  Unfortunately, the identification
of limited samples of object does not translate to detail constraints on 
the space density, scale height and luminosity function of the known classes of 
object and whilst our knowledge remains very sketchy, we are unable to give firm answers 
to many important astrophysical questions.

One important unanswered question is what fraction of the hard-band sources might be 
extragalactic interlopers seen through the high absorption column of the Galactic 
plane?  Since we require the X-ray sources in our sample to 
be pointlike (see \S3), this presumably mitigates against the presence of a significant 
number of clusters of galaxies, leaving active galactic nuclei (AGN) as the 
dominant extragalactic 
contaminant. From the investigation of the column densities in \S\ref{sec:nh},
it is clear that AGN would generally have HR~$>$~0.8. 
In an analysis of the 2--10 keV log N - log S relation measured in the Galactic 
plane based on the combination of {\it ASCA}, {\it Chandra} and {\it XMM-Newton}
observations, \citet{Hands04} demonstrate that at the flux levels
encompassed in our current survey, AGN might contribute as much 
as 50\% of the hard source population, with an even higher fraction likely
at fainter levels (\citealt{Ebisawa05}).   
Interestingly this rather closely matches the fraction of hard sources with HR~$>$~0.8.
However, we stress that these estimates are very sensitive to
the factor assumed for the hard-band transmission through the Galactic plane and
will vary significantly across our survey region; for example, \citet{Hands04} 
assume a transmission factor of 0.68 appropriate to (l,b)=(20,0).  
Further consideration of how the spatial density of the sources in the current 
survey varies as a function of flux and position, and the impact of AGN contamination,
will be the subject of a subsequent paper (Paper III -Warwick et al. 
{\it in preparation}).


\section{Conclusions}
\label{sec:concl}

We have used the 2XMMi source lists pertaining to 116 {\it XMM-Newton} observations 
targeted at the inner quadrant of the Galactic Plane, to construct a sample of 
serendipitous Galactic X-ray sources. The main properties of the 2204 sources which
comprise the sample are:

\begin{enumerate}

\item  The bulk of the sources have {\it total} count rates in the range 2-100 pn ct ks$^{-1}$
(0.5-12 keV). In the soft (0.5-2 keV) band, 2 pn ct ks$^{-1}$ corresponds to $F_X = 4 \times$10$^{-15}$ erg 
s$^{-1}$ cm$^{-2}$ (0.5--2 keV), assuming an absorbed power-law spectrum with $\Gamma$ = 2.5 and 
$\textit{N}_H$ = 10$^{21}$ cm$^{-2}$.
The same count rate in the hard (2-12 keV) band equates to 
$F_X = 2.5 \times$10$^{-14}$ erg s$^{-1}$ cm$^{-2}$ (2--10 keV), in this case assuming 
$\Gamma$ = 1.0 and $\textit{N}_H$ = 3$\times$10$^{22}$ cm$^{-2}$.

\item  Using a characterisation based on whether the majority of the counts 
were recorded below or above 2 keV, the sample splits rather cleanly into 1227 soft 
sources and 977 hard sources.

\item  Both the broad-band hardness ratio (HR) distribution and the {\it Band Index} 
($BI$) plots reflect a wide spread of underlying source spectra. For the soft sources,
the X-ray spectra may be represented as either power-law continua with $\Gamma \sim 2.5$ 
or as thermal spectra with kT $\sim 0.5$ keV with $N_H$ ranging from
$10^{20-22} \rm~cm^{-2}$.  For the hard sources, a significantly harder continuum 
form is likely, \ie  $\Gamma \sim 1$, with $N_H = 10^{22-24} \rm~cm^{-2}$.
The sources with HR$>$0.8 have column densities commensurate with the total Galactic
line-of-sight value.

\item   A high fraction ($^{>}_{\sim}$90\%)  of the soft sources have 
potential NIR (2MASS and/or UKIDSS) counterparts inside 
their error circles, consistent with the dominant soft X-ray source 
population being relatively nearby coronally-active stars. 
In contrast, the success rate in finding likely NIR counterparts 
to the hard X-ray sample is no more than $\approx25\%$ down to J=20,
set against a much higher chance coincidence rate. The make-up
of the hard band population in terms likely contributors such as
CVs, active binaries, relatively quiescent LMXBs/HMXB and other
classes of object remains uncertain.

\end{enumerate}

In future papers we will explore the average X-ray spectral 
properties of a subset of relatively bright sources drawn from our source sample 
(Paper II), the log N - log S curves for both the soft and hard source samples
(Paper III) and the broad-band colours of likely counterparts to the X-ray
sources (Paper IV). 

\section*{Acknowledgments}

This publication refers to data products from the Two Micron All Sky Survey, 
which is a joint project of the University of Massachusetts and the Infrared 
Processing and Analysis Center/California Institute of Technology, funded by the 
National Aeronautics and Space Administration and the National Science Foundation. 
In carrying out this research, use has been made of Aladin, the VizieR catalogue
access tool and Simbad at CDS, Strasbourg, France.

\end{document}